\documentclass[man,floatsintext,a4paper,donotrepeattitle]{apa7} 

\usepackage{float}
\usepackage{amsmath,amssymb}
\usepackage{mathtools}
\usepackage{cancel}
\usepackage{longtable}
\usepackage{array}
\usepackage{pdflscape}
\usepackage[figuresright]{rotating}

\newcommand{\bzero}{{\mathbf 0}}

\newcommand{\be}{{\mathbf e}}

\newcommand{\bp}{{\mathbf p}}

\newcommand{\bu}{{\mathbf u}}

\newcommand{\by}{{\mathbf y}}

\newcommand{\bB}{{\mathbf B}}

\newcommand{\bG}{{\mathbf G}}
\newcommand{\bH}{{\mathbf H}}
\newcommand{\bI}{{\mathbf I}}
\newcommand{\bJ}{{\mathbf J}}

\newcommand{\bL}{{\mathbf L}}
\newcommand{\bM}{{\mathbf M}}

\newcommand{\bT}{{\mathbf T}}

\newcommand{\bfeta}{{\boldsymbol\eta}}

\newcommand{\btheta}{{\boldsymbol\theta}}

\newcommand{\blambda}{{\boldsymbol\lambda}}
\newcommand{\bmu}{{\boldsymbol\mu}}

\newcommand{\bpi}{{\boldsymbol\pi}}

\newcommand{\brho}{{\boldsymbol\rho}}

\newcommand{\btau}{{\boldsymbol\tau}}

\newcommand{\bDelta}{{\boldsymbol\Delta}}
\newcommand{\bTheta}{{\boldsymbol\Theta}}
\newcommand{\bLambda}{{\boldsymbol\Lambda}}
\newcommand{\bXi}{{\boldsymbol\Xi}}

\newcommand{\bSigma}{{\boldsymbol\Sigma}}

\newcommand{\bPsi}{{\boldsymbol\Psi}}
\newcommand{\bOmega}{{\boldsymbol\Omega}}
\DeclareMathOperator{\diag}{diag}

\DeclareMathOperator{\E}{E}

\DeclareMathOperator{\Cov}{Cov}

\DeclareMathOperator{\N}{N}

\newcommand{\iid}{\,\overset{\text{iid}}{\sim}\,}

\DeclareMathOperator{\limvar}{limvar}

\newcommand{\bbR}{\mathbb{R}}

\newcommand{\cS}{{\mathcal S}}

\newcommand{\pl}{\operatorname{\ell_P}}

\newcommand{\mlepl}{\hat\btheta_{\text{PL}}}
\newcommand{\pimodpl}{\pi_{c_ic_j}^{(ij)}(\btheta)}
\newcommand{\ppl}{p_{c_ic_j}^{(ij)}}

\newcommand{\npl}{n_{c_ic_j}^{(ij)}}

\newcommand{\bBij}{{\bB_{c_{i}c_{j}}^{(ij)}}}

\newcommand{\bftheta}{\mbox{{\boldmath $\theta$}}}

\newcommand{\bfpi}{\mbox{{\boldmath $\pi$}}}

\usepackage{bm}
\usepackage{soul}
\usepackage{xcolor}
\usepackage{setspace}
\usepackage[style=apa,sortcites=true,sorting=nyt,backend=biber]{biblatex}
\title{Pairwise likelihood estimation and limited-information goodness-of-fit test statistics for binary factor analysis models under complex survey sampling}

\authorsnames[{1,2},2,2]{Haziq Jamil, Irini Moustaki, Chris Skinner\textsuperscript{\textdagger ,}}
\authorsaffiliations{{Universiti Brunei Darussalam},{London School of Economics and Political Science}}

\shorttitle{Estimation and testing under complex sampling design}

\authornote{%

\addORCIDlink{Haziq Jamil}{0000-0003-3298-1010}.

\addORCIDlink{Irini Moustaki}{0000-0001-8371-1251}.

\addORCIDlink{Chris Skinner}{0000-0002-6438-5531}.

The authors made the following contributions.
Haziq Jamil: Investigation, Methodology, Software, Visualization, Writing -- Review \& Editing;
Irini Moustaki: Conceptualization, Methodology, Formal Analysis, Validation, Writing -- Original Draft Preparation;
Chris Skinner: Methodology, Writing -- Review \& Editing.

Correspondence concerning this article should be addressed to Haziq Jamil, Faculty of Science, Universiti Brunei Darussalam, Jalan Tungku Link, Bandar Seri Begawan, Gadong BE1410. E-mail: \href{mailto:haziq.jamil@ubd.edu.bn}{haziq.jamil@ubd.edu.bn}.

\textsuperscript{\textdagger}Professor Chris Skinner passed away on the 21st of February 2020. He had been involved in the first stages of the current project, contributing to the methodology presented in the paper.

}

\abstract{This paper discusses estimation and limited-information goodness-of-fit test statistics in factor models for binary data using pairwise likelihood estimation and sampling weights. The paper extends the applicability of pairwise likelihood estimation for factor models with binary data to accommodate complex sampling designs. Additionally, it introduces two key limited-information test statistics: the Pearson chi-squared test and the Wald test. To enhance computational efficiency, the paper introduces modifications to both test statistics. The performance of the estimation and the proposed test statistics under simple random sampling and unequal probability sampling is evaluated using simulated data.}

\begin{document}
\maketitle

\setcounter{secnumdepth}{3}
\section{Introduction}

Latent variable models, such as factor analysis, are widely used in social sciences to measure abilities, attitudes, and behavior through observed categorical (binary, ordinal) or metric variables, also known as items or indicators. There are two main approaches for modeling categorical observed variables with latent variables, namely the complete information maximum likelihood approach (FIML) used in item response theory.
(see e.g. \cite{skrondal.rabe:04, bartholomew.ea:11}) and the limited-information approach used  in structural equation modelling (SEM) (e.g.
\cite{joreskog:90,muthen:84}). The latter uses first- and second-order statistics in the univariate and bivariate likelihood functions. The limited-information approach and, in particular, pairwise maximum likelihood (PML) is adopted here (\cite{joreskog.moustaki:01,liu:07,katsikatsou.ea:12,
katsikatsouthesis,xithesis}). PML is a special case of composite maximum likelihood (CML) estimation
(\cite{lindsay:88,varin:08,varin.ea:11}). CML estimators have the desired properties of being asymptotically consistent and normally distributed. Specifically, in the case of latent variable models, PML maximizes the sum of all bivariate log-likelihoods, which 
contains all the information needed for all model parameters. This approach offers a clear advantage over FIML estimation since it only requires evaluating up to two-dimensional integrals, each integrating over a different pair of underlying variables, irrespective of the number of observed or latent variables.

Pairwise likelihood ratio tests for overall goodness of fit,  nested models, and model selection criteria have been proposed by \textcite{katsikatsou.moustaki:16}.
However, due to data sparseness, overall goodness-of-fit (GOF) test statistics pose computational and theoretical challenges in high-dimensional settings.  
To address the sparseness issues, limited-information test statistics have been proposed. These test statistics are defined as quadratic forms of first- and second-order residuals and utilize information from the first- and second-order marginals (\cite{reiser:96,reiser:08, bartholomew.leung:02, maydeu.joe:05,maydeu.joe:06,cai.ea:06,cagnone.mignani:07}). \textcite{salomaa1990factor} also showed that the parameters of the factor model can be inferred from the first- and second-order marginal distributions of the observed variables. Pairwise likelihood estimation uses this property to estimate all model parameters from the bivariate likelihoods. Furthermore, lower-order residuals offer valuable evidence regarding model fit.
Previous research on limited-information tests mostly used a FIML estimator and simple random sampling.

In this paper, we first extend the PML to incorporate sampling weights, thus rendering it suitable for complex and informative sampling designs. Subsequently, we introduce a limited-information Pearson chi-squared test statistic alongside three variants of the Wald test statistic. These test statistics are designed to apply to simple random and complex sampling designs with unequal selection probabilities.

In complex sampling designs, the pairwise likelihood components are appropriately weighted using the provided sampling weights (\cite{skinner:89}). \textcite{skinner:89} and \textcite{muthen.satorra:95} in the context of structural equation modelling made a distinction between an aggregated approach, where the definitions of the model parameters take no account of complex population features such as stratification or clustering used in sampling and a disaggregated approach in which the complex population features are taken into account in the specification of the model of interest, for example, via fixed or random effect terms. Here, we focus on the aggregated approach. \textcite{skinner:19} provides an overview of categorical data analysis for complex surveys.

The paper is organized as follows: Section 2 presents the model framework. Section 3  gives the pairwise likelihood estimation for latent variable models for categorical responses under simple random and complex survey sampling. Section 4 proposes limited-information goodness-of-fit tests for complex sampling designs. Section 5 presents the results of the simulation studies, and section 6 concludes.

\section{Factor models for binary data}
\label{model}
The basic idea of latent variable analysis is as follows:
For a given set of response variables $y_1, \ldots, y_p$, one wants to find a set of latent factors $\eta_{1}, \ldots, \eta_{q}$, fewer in number than the observed variables that contain essentially the same information.
The latent factors are supposed to account for the dependencies of the response variables in the sense that if the factors are held fixed, the observed variables would be independent.
If both the response variables and the latent factors are normally distributed with zero means and unit variances, this leads to  the model \parencite{joreskog:79}
\begin{gather}
    E(y_i \,|\, \eta_1,  \ldots, \eta_q) = \lambda_{i1}\eta_1 +  \cdots + \lambda_{iq}\eta_q \label{eq:classicalfa} \\
    E(y_i y_j \,|\, \eta_1, \ldots, \eta_q) = 0, \ i \neq j.
\end{gather}
If the factors are independent, it follows that the correlation between $y_i$ and $y_j$ is $\sum_{l=1}^q \lambda_{il}\lambda_{jl}$.

Equation \eqref{eq:classicalfa} is a suitable representation of the factor analysis model if the response variables $y_i$ are continuous variables measured on an interval or ratio scale.
However, it cannot be used if the response variables $y_i$ are categorical.
In those cases, one must instead specify the probability of each response pattern as a function of $\eta_{1}, \ldots, \eta_{q}$:
\begin{equation}\label{eq:resppattern}
\pi = \Pr(y_1=c_1, \ldots, y_p=c_p \mid \eta_1, \ldots, \eta_q) =
f(\eta_1, \ldots, \eta_q),
\end{equation}
where $c_1,\ldots, c_p$ represent the different response categories of $y_1,\ldots, y_p$, respectively.
In this paper, we consider the case of binary response variables ($c_i \in \{0,1\}$) and continuous latent variables.
The methodology presented can easily be extended to ordinal variables.

Let $\mathbf{y}=\left(y_{1},\ldots,y_{p}\right)'$ denote the vector of $p$ binary observed variables.
Thus, there are $R = 2^p$ possible response patterns of the form $\mathbf{y}_{r}=\left( c_{1}, c_{2},\ldots, c_{p} \right)'$, $c_{i} \in \{0,1\}$. 
For a random sample $\cS = \{ \by^{(h)}\}_{h=1}^n$ of size $n$, where $\by^{(h)}$ denotes the value of $\by$ for unit $h$, the log-likelihood is
\begin{equation}
\log L(\boldsymbol{\theta} ; \cS )=\sum_{r=1}^{R} n_{r} \log \pi_{r}(\boldsymbol{\theta}), \label{eq:log_likel}
\end{equation}
where $n_r$ and $\pi_{r}(\boldsymbol{\theta})$ are the observed frequency and the probability under the model, respectively, for the response pattern $r$, for some parameter vector $\bftheta$. 
As such,  $\sum_{r=1}^R n_r = n$ and $\sum_{r=1}^R \pi_{r}(\boldsymbol{\theta}) = 1$.

In the case of complex survey designs where the sample is drawn from a finite population of size $N$, we maximize the weighted log-likelihood 
\begin{equation}\label{eq:cxlog_likel}
\log L(\boldsymbol{\theta}; \cS) = \sum_{r=1}^{R}\hat{n}_{r}\ln\pi_{r}(\boldsymbol{\theta}),
\end{equation}
where $\hat{n}_{r}$ is the sum of survey weights across sample units with response pattern $r$. 
If we denote the survey weight for unit $h$ in $\cS$ as $w_h$, we can write $\hat{n}_{r}=\sum_{h \in \cS} w_h [\mathbf{y}^{(h)} = \mathbf{y}_r]$.
Here, the notation $[\cdot]$ denotes the Iverson bracket: $[A] = 1$ if the proposition $A$ is true, and $[A]=0$ otherwise.
\textcite{skinner:89}  showed that the parameter estimates from the weighted log-likelihood (pseudo-likelihood) are consistent under any sampling scheme.

We adopt the underlying response variable approach (UV) to model the binary variables.
Each binary variable $y_i$ is taken to be a manifestation of an underlying continuous and normally distributed random variable $y_i^*$.
The connection between the observed $y_i$ and the underlying variable $y_i^*$ is as follows:
\begin{equation}
    y_i = \begin{cases}
    1 & y_i^* > \tau_i \\
    0 & y_i^* \leq \tau_i,
    \end{cases}
    \nonumber
\end{equation}
where $\tau_{i}$ is a threshold associated with the underlying variable $y_i^*$.
For convenience, the distribution of $y_{i}^*$ is assumed to be standard normal.

The factor model is of the form
\begin{equation}
\mathbf{y}^{\star}= \boldsymbol\Lambda\boldsymbol{\eta}+\boldsymbol{\epsilon}, \label{eq: FA model}
\end{equation}
where $\mathbf{y}^{\star}=(y_{1}^*,\cdots,y_{p}^*)'$ is the $p$-dimensional vector of the underlying variables, $\boldsymbol\Lambda$ is the $p\times q$ matrix of loadings, and $\boldsymbol{\epsilon}$ is the $p$-dimensional vector of unique
variables.
In addition, it is assumed that $\boldsymbol{\eta}\sim \N_{q}(\mathbf{0},\bPsi)$ where $\bPsi$ contains ones on its main diagonal.
In this way, $\bPsi$ is the correlation matrix of latent factors.
Further it is assumed $\boldsymbol{\epsilon}\sim \N_{p}(\mathbf{0},\bTheta_\epsilon)$ with $\bTheta_\epsilon$ a diagonal matrix defined by $\bTheta_{\epsilon}=\mathbf I - \diag(\bLambda\bPsi\bLambda')$, and that $\Cov(\epsilon_i, \eta_j) = 0$ for each combination of $i=1,\dots,p$ and $j=1,\dots,q$.

The parameter vector $\boldsymbol{\theta}' = \left(\boldsymbol{\lambda}', \boldsymbol{\rho}', \boldsymbol{\tau}'\right)$ contains $\boldsymbol{\lambda}$ and $\boldsymbol{\rho}$ which are the vectors of the free non-redundant parameters in matrices $\bLambda$ and $\bPsi$ respectively, as well as $\boldsymbol{\tau}$ which is the vector of all free thresholds.
Under the UV framework, the probability of a response pattern $r$ is written as
\begin{equation}
\pi_{r}(\boldsymbol{\theta})
= \Pr\left(y_{1}=c_{1},\ldots,y_{p}=c_{p}; \bftheta\right)
= \idotsint \phi_{p}(\mathbf{y}^{\star} \,|\, \mathbf 0 , \bSigma_{\mathbf{y}^{\star}}) \ \text{d} \mathbf{y}^{\star}, \label{eq:p_dimen_normal_prob}
\end{equation}
where $\phi_{p}(\cdot \,|\, \boldsymbol\mu, \bSigma)$ is the $p$-dimensional Gaussian probability density function with mean vector $\bmu$ and variance-covariance matrix $\bSigma$.
The variance-covariance matrix of the underlying variables is the correlation matrix
$\bSigma_{\mathbf{y}^{\star}}=\bLambda\bPsi\bLambda'+\bTheta_\epsilon$.
The $p$-dimensional integral \eqref{eq:p_dimen_normal_prob} is evaluated on the domain of $\mathbf y^*$ constrained accordingly using the thresholds $\tau_i$ so that the manifest variables correspond to $\mathbf c = (c_1,\dots,c_p)$.


\section{Pairwise likelihood estimation}

To mazimize the log-likelihood function in (\ref{eq:log_likel})
over the parameter vector $\boldsymbol{\theta}$, we need to evaluate
 \eqref{eq:p_dimen_normal_prob}, and that cannot be expressed in a closed form.
As a result, limited-information estimation methods have been developed and are now widely available in commercial software. The most popular method is the three-stage weighted least squares estimation  \parencite{joreskog:90, joreskog:94, muthen:84}.
However, we will be using a different method known as pairwise maximum likelihood as proposed in \textcite{katsikatsou.ea:12} and implemented in the R Package \texttt{lavaan} (\cite{Rosseel2012_JSS}).  The method is briefly explained below.

To define the pairwise likelihood, we assume that $\pimodpl$ is the probability that both $y_i$ and $y_j$ take on values $c_i$ and $c_j$, respectively, under a particular model.
For a random sample of size $n$, the pairwise log-likelihood is
\begin{equation}\label{eq:PL}
\sum_{i<j} \sum_{c_i=0,1} \sum_{c_j=0,1} \npl \log \pimodpl,
\end{equation}
where $\npl$ is the observed frequency of sample units with $y_i=c_i$ and $y_j=c_j$.
To account for complex sampling, we use the weighted pairwise log-likelihood instead:
\begin{equation}\label{eq:PLwt}
\pl(\btheta; \cS) = \sum_{i<j} \sum_{c_i=0,1} \sum_{c_j=0,1} \ppl \log \pimodpl,
\end{equation}
where
$\ppl =  \sum_{h \in \cS} w_h [y_i^{(h)} = c_i, y_j^{(h)} = c_j] \Big/ \sum_{h \in s} w_h,$ and $y_i^{(h)}$ refers to the $i$th item for sampling unit $h$ in $\cS$.
We have introduced survey weights $w_h$ and scaled them by the constant factor $1 / \sum_{h \in s} w_h$.
The pairwise likelihood only requires the calculation of up to two-dimensional normal probabilities, regardless of the number of observed or latent variables.
In this way, it is always computationally feasible.

The maximum pairwise likelihood estimator (MPLE) $\mlepl$ solves the estimating equations $\bbR^m \ni \nabla \pl(\btheta; \cS) = \bzero$, where the entries of this score vector are
\begin{equation} \label{eq:bivderall}
\frac{\partial\!\pl(\btheta; \cS)}{\partial\theta_k} =
\sum_{i<j} \sum_{c_i=0,1} \sum_{c_j=0,1}
\frac{\ppl}{\pimodpl} \frac{\partial \pimodpl}{\partial\theta_k}, k=1,\dots,m.
\end{equation}
The $(k,l)$ element of the $m\times m$ Hessian matrix, $\nabla^2 \pl(\btheta; {\bf y})$, is
\begin{equation}\label{eq:secderpl}
\frac{\partial\!\pl(\btheta; \cS)}{\partial\theta_k\partial\theta_l}
=
\sum_{i<j} \sum_{c_i=0,1} \sum_{c_j=0,1}
\frac{\ppl}{\pimodpl}
\left\{
\frac{\partial^2\pimodpl}{\partial\theta_k\partial\theta_l}
-
\frac{1}{\pimodpl}
\frac{\partial\pimodpl}{\partial\theta_k}
\frac{\partial\pimodpl}{\partial\theta_l}
\right\}.
\end{equation}
Using Taylor expansion, we may write
\begin{equation}\label{eq:PL2}
\mlepl - \btheta =
 \{   \nabla^2 \pl(\bftheta; \cS)\}^{-1}
\nabla\pl(\btheta; \cS) + o_p(n^{-1/2}).
\end{equation}
It follows that
\begin{equation}
\sqrt{ n} \big(\mlepl - \btheta\big) \xrightarrow{\text d} \N_m \big(\bzero, \bH(\btheta)^{-1}\bJ(\btheta)\bH(\btheta)^{-1}\big),
\end{equation}
where $m$ is the dimension of $\bftheta$, $\bH(\btheta)=\operatorname{E}\{-\nabla^2 \pl (\btheta;{\bf y})\}$ is the sensitivity matrix, and $\bJ(\btheta) = \operatorname{Var}\left\{ \nabla \pl(\boldsymbol{\theta};\mathbf{y})\right\}$ is the variability matrix. 
$\bG = \bH(\boldsymbol{\theta})\bJ(\boldsymbol{\theta})^{-1}\bH(\boldsymbol{\theta})$ is known as the Godambe information (\cite{godambe:60}) or sandwich information matrix and it indicates the loss of efficiency compared to full information maximum likelihood estimation.
If $\pl$ is a true log-likelihood function, then the Godambe, sensitivity, and Fisher information matrix are all identical.
In practice, we can estimate the $\bH$ and $\bJ$ matrices using 
(\cite{varin.ea:11,zhao.joe05})
\begin{gather}
\hat\bH (\mlepl)= -\frac{1}{\sum_{h\in\cS} w_h} \sum_{h\in\cS} \nabla^2\pl(\btheta; \by^{(h)})
\Big|_{\btheta=\mlepl} \\
\text{and} \nonumber \\
\hat\bJ (\mlepl)=  \frac{1}{\sum_{h\in\cS} w_h} \sum_{h\in\cS} \nabla\pl(\btheta; \by_h)\nabla\pl(\btheta; \by^{(h)})' \Big|_{\btheta=\mlepl}.
\end{gather}

The estimator for the variability matrix $\bJ$ assumes independent observations. However, in cluster sampling, the observations are no longer independent. The variance estimator for the weighted pairwise estimates should be adjusted  to account for the dependencies and stratification (\cite[see][]{binder:83,wolter:07,asparouhov2005sampling}).
This gives
\begin{equation}\label{eq:clusterJ}
\hat\bJ (\mlepl)=\frac{1}{n} \sum_{a}\frac{n_a}{n_a-1} \sum_{b} (z_{ab}-\bar{z_a})(z_{ab}-\bar{z_a})',
\end{equation}
where $n_a$ is the number of sampled clusters from stratum $a$, $z_{ab}=\sum_{h\in\cS_{ab}} \nabla \pl(\btheta;\by^{(h)})$ is the total value of the score for all individuals $h$ in cluster $b$ in stratum $a$, and $\bar{z_a}$ is the average of $z_{ab}$. 
In addition, \textcite{varin:08} discusses alternative empirical estimators for the variability matrix in the presence of clustering in a composite likelihood framework.
\section{Goodness-of-fit}
\label{gof}

The overall goodness of fit of the model can be tested by constructing a likelihood ratio test using the full log-likelihood given in equation \eqref{eq:cxlog_likel} for testing $H_{0}: \pi_{r} = \pi_{r}(\bftheta)$ against $H_1: \pi_{r} \neq \pi_{r}(\bftheta)$, for all $r=1,\dots,R$, where $\btheta$ is a vector of all free parameters and $r$ runs over all possible response patterns.

Let $\hat\pi_r = \pi_r(\hat{\btheta})$, where $\hat{\btheta}$ denotes the FIML estimates, and let $\hat p_r = n_r/n$ denote the sample proportion of response pattern $r$.
The maximum log-likelihood value under $H_0$ is $n\sum_{r=1}^R p_r \log \hat\pi_r$, while the maximum log-likelihood value under $H_1$ is $n\sum_{r=1}^R p_r \log p_r$.
The likelihood ratio (LR) test statistic is:
\begin{equation}\label{eq:lrchisq}
X_{\text{LR}}^2
= 2n \sum_{r=1}^R  p_{r} \log ( p_r / \hat \pi_r).
\end{equation}
Under $H_{0}$, \eqref{eq:lrchisq} is distributed approximately as $\chi^{2}$
with degrees of freedom equal to the number of independent response patterns
minus one minus the number of elements in $\bftheta$. For the $q$-factor model, the degrees of
freedom is
$2^p - p(q+1)-1$.
Alternatively, one can use the Pearson goodness-of-fit (GF) test statistic written as:
\begin{equation}
X_{\text{GF}}^2
= n \sum_{r=1}^R( p_{r}-\hat{\pi}_{r})^2/\hat{\pi}_{r}\;.
\label{eq:gfchisq} \end{equation}
Under $H_{0}$, both test statistics (\ref{eq:lrchisq} and \ref{eq:gfchisq}) have the same asymptotic distribution.

In principle, these tests are possible to use with FIML. They cannot be used
with the UV approach because this does not maximize an overall likelihood function, so
the $\hat{\pi}_{r}$ are not computed.
In practice, when dealing with sparse data characterised by zero or small frequencies $n_r$, the approximation to the chi-square distribution may fail and lead to unreliable results (\cite[see e.g.][]{reiser.vandenberg:94}).

To address the sparseness issue, limited-information test statistics focus on how well the model fits first- and second-order marginals (marginal proportions) such as univariate, bivariate, and trivariate rather than the entire response pattern.
In the following sections, we introduce limited-information chi-square test statistics for evaluating the adequacy of a parametric model under simple random sampling and complex sampling designs and pairwise likelihood estimation.

\subsection{Definitions}

Define $\dot{\pi}_i= \Pr(y_i = 1)$ to be the first-order marginal proportion of a positive response for the $i$th binary variable $i=1,\dots,p$, and let $\dot{\bfpi}_1 = (\dot\pi_1,\dots,\dot\pi_p)'$ be the $p\times 1$ vector containing all such first-order moments.
Further define with $\dot\pi_{ij} = \Pr(y_i=1,y_j=1)$ the second-order marginal proportion for a pair of variables $i,j=1,\dots,p$, and collect these proportions into the $\binom{p}{2}$-vector $\dot{\bpi}_2 = \left( \dot\pi_{ij} \right)_{i<j}$.
Finally, let
${\bpi}_2 = \begin{psmallmatrix} \dot{\bfpi}_1 \\ \dot{\bfpi}_2 \end{psmallmatrix}$
be the vector that contains both these first- and second-order marginal proportions with dimension
$S = p + \binom{p}{2} = p(p+1)/2$.

The vectors $\dot{\bfpi}_1$ and $\dot{\bfpi}_2$ are formed by summing the joint proportions of the response patterns $\bpi$ that satisfy specific criteria.
For example, the first-order moment $\dot{\pi}_i$ for a specific variable $i$ is obtained by summing the joint proportions $\pi_r$ where the response pattern $r$ takes the value 1 in the $i$th position.
Similarly, the second-order moment $\dot{\pi}_{ij}$ for a pair of variables $i$ and $j$ is obtained by summing the values of $\pi_r$ for which the response patterns $r$ take on a value of 1 at both locations $i$ and $j$ simultaneously.
In other words, there exists a transformation matrix $\bT_2: \mathbb{R}^R \to \mathbb{R}^S$ that maps the joint proportions $\bpi$ to the transformed vectors $\bpi_2$, i.e., $\bpi_2 = \bT_2 \bpi$.
This matrix consists of 0s and 1s and has a rank of $S$, where $S$ represents the dimension of the transformed space.
A more detailed treatment of this transformation matrix can be found in \textcite{reiser:96} or \textcite{maydeu.joe:05}

Let ${\bf p}$ denote the $R \times 1$ vector of sample proportions based on a sample of size $n$, which estimates the population proportions vector $\bpi$.
Assuming independent and identically distributed observations, it is known that
\begin{equation}
\sqrt{n}({\bf p} - \bfpi) \xrightarrow{\text d} \N_R(\bzero, \bSigma),
\label{eq0}
\end{equation}
where $\bSigma$ is the multinomial covariance matrix given by $\bSigma = \diag(\bfpi) - \bfpi\bfpi'$.
In the case of a complex sampling design, the vector $\bp$ becomes the weighted vector of proportions with elements $\sum_{h \in \cS} w_h [\mathbf{y}^{(h)} = \mathbf{y}_r] \big/  \sum_{h \in \cS} w_h$.
The vector $\bfpi$  may now be either the vector of proportions in the finite population or a superpopulation.
Under suitable conditions (e.g., \cite{fuller:09} sec. 1.3.2), we still have a central limit theorem as in \eqref{eq0},
where the covariance matrix $\bSigma$ does not take a multinomial form.
In either case, the central limit theorem in \eqref{eq0} readily transforms to the limited-information case.
Consider the $S\times 1$ vectors $\bfpi_2=\bT_2\bfpi$ and ${\bf p}_2= \bT_2{\bf p}$, whose meanings are self apparent.
Then
\begin{equation}
\sqrt{n}(\bp_2 - \bpi_2) \xrightarrow{\text d} \N_S(\bzero,\bSigma_2),
\label{eq00}
\end{equation}
where $\bSigma_2 = \bT_2 \bSigma \bT_2'$.
Because $\bT_2$ is of full column rank $S$, $\bSigma_2$ is also of full rank $S$.

\subsection{Test statistics for simple and composite hypotheses}
In the case of a simple null hypothesis $H_0:{\bfpi}={\bfpi}_0$ and independent and identically distributed observations, \textcite{maydeu.joe:05} proposed the test statistic:
\begin{equation}
X^2 = n(\bp_2-\bpi_{20})'\bSigma^{-1}_{20}(\bp_2-\bpi_{20}),
\end{equation}
where $\bpi_{20}$ and $\bSigma_{20}$ are the values of $\bpi_2$ and  $\diag(\bpi_2)-\bpi_2\bpi_2'$, respectively, when ${\bfpi}={\bfpi}_0$.
As $n$ increases, the test statistic tends towards a $\chi^2$ distribution with $S$ degrees of freedom.

To evaluate the adequacy of the parametric model (composite hypothesis),  we define the null hypothesis $H_0: \bfpi=\bfpi({\bftheta})$ for some $\bftheta$ against an alternative hypothesis  $H_1:\bfpi\neq \bfpi({\bftheta})$ for any $\bftheta$.
To construct the test statistics, we consider the residual vector $\hat{{\be}}_2={\bp}_2-\bfpi_2(\mlepl)$, where $\bp_2$ represent the first- and second-order observed proportions while $\bpi_2(\mlepl)$ represent the same predicted by the parametric model when evaluated at pairwise maximum likelihood estimate $\hat{\bftheta}_{\text{PL}}$.

We first derive the asymptotic distribution of  $\hat{{\bf e}}_2$.
Noting that $\bpi_2(\bftheta) = \bT_{2}\bpi(\btheta)$, a Taylor series expansion about the MPLE gives
\begin{equation}
\bpi_2(\mlepl)
= \bpi_2(\btheta) + \bT_2 \bDelta (\mlepl - \btheta) + o_p(|\mlepl - \btheta |),
\end{equation}
where $\bDelta=\frac{\partial{\bfpi(\bftheta)}}{\partial{\bftheta}}$.
Let $\bDelta_2$ be defined as $\bDelta_2= \frac{\partial{\bfpi_2(\bftheta)}}{\partial{\bftheta}} = \bT_2 \bDelta$.
Using the earlier Taylor expansion of the score vector in \eqref{eq:PL2}, we have that
\begin{align}
\hat{\be}_2
&= {\bp}_2 - \bpi_2(\btheta) - \bDelta_2
\{ \nabla^2 \pl (\btheta) \}^{-1}
\nabla \pl(\btheta)
+ o_p(|\mlepl - \btheta |).
\label{eq:RES}
\end{align}

In order to express $\nabla \pl(\boldsymbol{\theta})$ in terms of ${\bp}_2-\bpi_2(\btheta)$, we use
\begin{align}
\sum_{i<j} \sum_{c_i=0,1}\sum_{c_j=0,1} \, &  \frac{1}{\pimodpl}  \frac{\partial \pimodpl}{\partial\theta_k} \big( p_{c_ic_j}^{(ij)} -  \pimodpl \big) \label{eq:scorease2} \\
&= \sum_{i<j} \sum_{c_i=0,1}\sum_{c_j=0,1}
\frac{\ppl} {\pimodpl} \frac{\partial \pimodpl}{\partial\theta_k}
-
\cancel{\sum_{i<j} \sum_{c_i=0,1}\sum_{c_j=0,1} \frac{\partial \pimodpl}{\partial\theta_k}}  \nonumber \\
&=  \frac{\partial\!\pl(\btheta)}{\partial\theta_k}, \nonumber
\end{align}
as a means of expressing the $k$th component of the score vector in \eqref{eq:bivderall}.
The cancellation in the last line above follows from the fact that for all $k=1,\dots,m$ and any pairwise proportions with index $i,j \in \{1,\dots,p\}$, $i<j$,
\begin{equation}\label{eq:sumderzero2}
\sum_{c_i=0,1}\sum_{c_j=0,1} \pimodpl = 1 \ \ \Rightarrow \ \  \sum_{c_i=0,1}\sum_{c_j=0,1} \frac{\partial \pimodpl}{\partial\theta_k}  = 0.
\end{equation}
Furthermore, for any pair $(i,j)$ we require an $1 \times S$ design vector $\bB_{c_{i}c_{j}}^{(ij)}$ such that
\begin{equation}\label{eqder1}
p_{c_{i}c_{j}}^{(ij)}- \pimodpl
= \bB_{c_{i}c_{j}}^{(ij)} \big(\bp_2- \bpi_2(\btheta)\big),
\end{equation}
since the pairwise proportions may be obtained through a linear combination of the first- and second-order proportions.
By substituting \eqref{eqder1} into \eqref{eq:scorease2} and amalgamating the preceding terms in \eqref{eq:scorease2} with the design vector $\bB_{c_{i}c_{j}}^{(ij)}$,
we find an $m \times S$ matrix $\bB(\bftheta)$ that depends on the parameters $\btheta$,  that satisfies
\begin{equation}\label{eqder2}
\nabla \pl(\btheta) = \bB(\btheta)(\bp_2-\bpi_2(\btheta)).
\end{equation}
The intermediate steps from equation \eqref{eqder1} to equation \eqref{eqder2} are given in Appendix \ref{apx:proof}.
Hence, from (\ref{eq:RES}), and multiplying through by $\sqrt n$, we get
\begin{equation}
\sqrt n \, \hat\be_2 =
\left(\bI - \bDelta_2 \{ \nabla^2 \pl(\btheta) \}^{-1} \bB(\btheta)\right) \cdot \sqrt n (\bp_2 - \bpi_2(\btheta))
+o_p(1).
\end{equation}
So from (\ref{eq00}), the consistency of the Hessian, and using Slutzky's theorem, we have the limiting distribution of the lower order residuals under $H_0$.
It is $\sqrt{n} \, \hat\be_2  \xrightarrow{\text d} \N_S(\bzero, \bOmega_2)$, where
\begin{equation}\label{eq:Omega2}
\bOmega_2 =
\left(\bI - \bDelta_2 \bH(\btheta)^{-1} \bB(\btheta) \right)
\bSigma_2
\left(\bI - \bDelta_2 \bH(\btheta)^{-1} \bB(\btheta) \right)'.
\end{equation}

To calculate test statistics, it is necessary to use an estimator for the asymptotic covariance matrix of $\hat\be_2$.
Evaluate $\frac{\partial{\bpi(\btheta)}}{\partial{\btheta}}$ at the PL estimate
$\mlepl$ to obtain $\hat{\bDelta}_2 = \bT_2\bDelta |_{\btheta = \mlepl}$, and set
\begin{equation}
\hat{\bOmega}_2 =
\left(\bI - \hat{\bDelta}_2 \hat\bH(\mlepl)^{-1} \, \bB(\mlepl) \right)
\hat{\bSigma}_2
\left(\bI - \hat{\bDelta}_2 \hat\bH(\mlepl)^{-1} \, \bB(\mlepl) \right)'
\end{equation}
with $\hat{\bSigma}_2=\bT_2\hat{\bSigma}\bT_2'$.
The construction of the estimator $\hat{\bSigma}$ is discussed in Section \ref{sec:estSigma}.
In the case of iid observations with a multinomial covariance matrix, we may set $\hat{\bSigma} = \diag(\bpi(\mlepl)) - \bpi(\mlepl) \bpi(\mlepl)'$.

Subsequently, the aim is to build limited-information goodness-of-fit test statistics taking the quadratic form
\begin{equation}\label{eq:quadformtest}
X^2 = n \hat{\be}_2' \hat{\bXi} \hat{\be}_2
\end{equation}
such that $\hat{\bXi} \xrightarrow{\text P} \bXi$ as $n\to\infty$ where $\bXi$ is some $S \times S$ weight matrix.
Generally, $p$-values relating to $X^2$ will be compared against a reference chi-square distribution.
This is because the quadratic form in \eqref{eq:quadformtest} converges in distribution to $\sum_{s=1}^S \delta_s U_s$ as $n\to\infty$, where each $U_s\iid\chi^2_1$ and $\delta_1,\dots,\delta_S$ are the eigenvalues of $\bOmega_2^{1/2}\bXi\bOmega_2^{1/2}$ (or of $\bL'\bXi\bL$, where $\bL$ is the lower triangular matrix from the Choleski decomposition of $\bOmega_2$).
An exact asymptotic chi-square distribution arises if $\bXi$ is chosen such that the eigenvalues $\delta_s$ are either 0 or 1.
If not, a sum of scaled chi-square variates arises.
A chi-square distribution can approximate this mixture of chi-square variates, but its degrees of freedom must be estimated.
We describe in Appendix \ref{sec:estchisq} the moment matching procedure to estimate the degrees of freedom of $X^2$ in such cases.

In what follows, we discuss Wald-type, Pearson, and other test statistics that utilise different forms for the weight matrix $\bXi$.

\subsubsection{Wald-type tests}
A Wald test statistic can be constructed by selecting the inverse of ${\bOmega}_2$ as the weight matrix $\bXi$.
This choice ensures that the resulting test statistic follows an exact asymptotic chi-square distribution under the null hypothesis $H_0$.
Due to numerical instabilities, the rank of $\bOmega_2$ may be deficient, causing problems for calculating the inverse.
In our implementation, we adopt the Moore-Penrose inverse $\hat{\bXi} = \hat{\bOmega}_2^+$, as done in \textcite{reiser:96}. Under the null hypothesis ($H_0$), this test statistic is asymptotically chi-squared distributed with degrees of freedom equal to the rank of $\hat{\bOmega}_2$, which falls between $S-m$ and $S$ (\cite{maydeu.joe:05}). Here $m$ represents all the free model parameters. One drawback is that the degrees of freedom are initially unknown and can only be estimated after estimating $\hat{\bOmega}_2$ by inspecting the magnitude of its eigenvalues. As a result, the $p$-value depends on the assessment of which eigenvalues are greater than zero. Additionally, inverting $\hat{\bOmega}_2$ can be computationally challenging when its dimensions are large.

A different version of the Wald test is also considered, referred to as the \emph{diagonal Wald test}, where $\hat{\boldsymbol\Xi} = \operatorname{diag}(\hat{\boldsymbol\Omega}_2)^{-1}$ is used instead of the pseudo inverse of ${\boldsymbol\Omega}_2$.
The motivation behind this choice is enhanced numerical stability and computational simplicity. Inverting a diagonal matrix is straightforward compared to inverting a full matrix, especially as the number of items $p$ increases.
However, the diagonal Wald test statistic converges to a sum of scaled chi-squares as mentioned above and, in finite samples, can be approximated by an appropriate chi-square distribution.

The drawbacks of ${\bOmega}_2$ disscused above  led to  \textcite{maydeu.joe:05,maydeu.joe:06} suggested using a weight matrix \({\boldsymbol\Xi}\) such that \({\boldsymbol\Omega}_2\) is a generalised inverse of \({\boldsymbol\Xi}\), i.e.~\({\boldsymbol\Xi}= {\boldsymbol\Xi}{\boldsymbol\Omega}_2{\boldsymbol\Xi}\).
Let \({\boldsymbol\Delta}_2^\perp\) be an \(S \times (S-m)\) orthogonal complement to \({\boldsymbol\Delta}_2\), i.e.~it satisfies \(({\boldsymbol\Delta}_2^\perp)'{\boldsymbol\Delta}_2 = {\mathbf 0}\).
Then due to the asymptotic normality of $\hat{\be}_2$ we see that as $n\to \infty$,
\begin{equation}\label{eq:e2OC}
\sqrt n ({\boldsymbol\Delta}_2^\perp)' \hat {\mathbf e}_2
\xrightarrow{\text d}
\N_{S-m} (\mathbf 0, ({\boldsymbol\Delta}_2^\perp)' {\boldsymbol\Omega}_2 {\boldsymbol\Delta}_2^\perp).
\end{equation}
Because of \eqref{eq:Omega2}, the asymptotic covariance matrix may be written
\begin{equation}
({\boldsymbol\Delta}_2^\perp)' {\boldsymbol\Omega}_2 {\boldsymbol\Delta}_2^\perp = ({\boldsymbol\Delta}_2^\perp)' {\boldsymbol\Sigma}_2 {\boldsymbol\Delta}_2^\perp,
\nonumber
\end{equation}
since all the multiplications of \({\boldsymbol\Delta}_2\) with its orthogonal complement cancels out.
By letting
\begin{equation}
{\boldsymbol\Xi}= {\boldsymbol\Delta}_2^\perp \big( ({\boldsymbol\Delta}_2^\perp)' {\boldsymbol\Sigma}_2 {\boldsymbol\Delta}_2^\perp \big)^{-1} ({\boldsymbol\Delta}_2^\perp)',
\nonumber
\end{equation}
we can then verify \({\boldsymbol\Xi}= {\boldsymbol\Xi}{\boldsymbol\Omega}_2{\boldsymbol\Xi}\); that is, \({\boldsymbol\Omega}_2\) is a generalised inverse of \({\boldsymbol\Xi}\).
Let \(\hat{\boldsymbol\Xi}\) be an appropriate estimate of \({\boldsymbol\Xi}\), e.g.~by replacing the matrices above with their corresponding hat versions.
The implication here is that
\begin{equation}
X^2
= n \, \hat{\mathbf e}_2' \hat{\boldsymbol\Xi}\hat{\mathbf e}_2
= n \,\hat{\mathbf e}_2'  \hat{\boldsymbol\Delta}_2^\perp \big( (\hat{\boldsymbol\Delta}_2^\perp)' \hat{\boldsymbol\Sigma}_2 \hat{\boldsymbol\Delta}_2^\perp \big)^{-1} (\hat{\boldsymbol\Delta}_2^\perp)' \hat{\mathbf e}_2
\end{equation}
converges in distribution to a \(\chi^2_{S-m}\) variate as \(n\to\infty\) due to \eqref{eq:e2OC} and Slutsky's theorem.
The degrees of freedom are as such because \({\boldsymbol\Delta}_2^\perp\) is of full column rank \(S-m\) and hence \({\boldsymbol\Xi}\) is also of rank \(S-m\), since the dimensions of a vector space and its orthogonal complement always add up to the dimension of the whole space.
We refer to this test as the \emph{variance-covariance free (VCF) Wald test}.

\subsubsection{Pearson test statistic}

To construct a Pearson test statistic, let \(\hat{\boldsymbol\Xi}^{-1} = \hat{\mathbf D}_2 = \mathop{\mathrm{diag}}( \bpi_2(\mlepl) ) \) which converges in probability to
\({\mathbf D}_2= \mathop{\mathrm{diag}} ({\boldsymbol\pi}_2({\boldsymbol\theta})) \) as $n\to\infty$.
Then we can see that
\begin{equation}\label{eq:Pearsontest}
X^2 = n \, \hat{\mathbf e}_2'\hat{\mathbf D}_2^{-1}\hat{\mathbf e}_2 = n \sum_{r=1}^R \frac{(\dot p_r - \dot\pi_r(\hat{\boldsymbol\theta}))^2}{\dot\pi_r(\hat{\boldsymbol\theta})} +
n \sum_{r<s} \frac{(\dot p_{rs} - \dot\pi_{rs}(\hat{\boldsymbol\theta}))^2}{\dot\pi_{rs}(\hat{\boldsymbol\theta})},
\end{equation}
which resembles the traditional Pearson chi-square test statistic.
Here, the $\dot p_r$ and $\dot p_{rs}$ are the first- and second-order  sample proportion estimates to $\dot\pi_r$ and $\dot\pi_{rs}$ respectively.
When computing the Pearson test statistic, there is no need to invert   $\hat{\bOmega}_2$ as in the Wald test. However, 
unlike the traditional Pearson test statistic, this does not follow  an asymptotic chi-square distribution because the independence requirement is violated in the second part of the above sum.
Instead, it converges to a sum of scaled chi-squares, whose distribution may be approximated by an appropriate chi-square distribution. Similarly, here, we employ the moment matching adjustment described in Appendix \ref{sec:estchisq} to approximate the distribution of the Pearson test statistic.

\subsubsection{Other test statistics}

Two alternative weight matrices $\hat\bXi$ are proposed and selected for their ease of computation in determining the test statistic $X^2$.
The first option is the identity matrix.
This results in a chi-square value resembling the sum of squared residuals (RSS), which gives a pure measure of the discrepancy between the fitted model and the data.
We refer to this as the RSS test.

The second option involves using multinomial weights $\hat\bXi^{-1} = \diag(\hat\bpi_2) - \hat\bpi_2\hat\bpi_2'$.
This matrix is a consistent estimator for the (inverse of) the multinomial covariance matrix $\bSigma_2$ for the univariate and bivariate proportions $\bp_2$.
\textcite{bartholomew.leung:02} had previously examined a variation of this test, focusing solely on the bivariate combinations and excluding the univariate ones.

In both choices mentioned here, the asymptotic distribution of the resulting test statistics is a sum of scaled chi-squares, so it is approximated by a chi-square whose degrees of freedom must be estimated using the data by employing the moment matching adjustment described in Appendix \ref{sec:estchisq}.

\subsection{Estimation of the covariance matrix under complex sampling}
\label{sec:estSigma}

We now consider how to construct the covariance estimator of the proportion,  $\hat{\bSigma}$ of $\bSigma$, in \eqref{eq0} for complex sampling designs.
We noted earlier that under iid assumptions, we may use the multinomial expression $\hat{\bSigma}=\diag(\bp)-\bp \bp'$ in the simple hypothesis case and $\hat{\bSigma}=\diag(\bpi(\mlepl))-\bpi(\mlepl) \bpi(\mlepl)'$ in the composite hypothesis case.
We now consider the case of complex sampling, in particular, stratified multistage sampling.

From (\ref{eq0}) we may write
\begin{equation}
\bSigma
= \limvar \{ \sqrt n (\bp - \bpi) \}
= \limvar \left\{ \sqrt n \left(  \frac{\sum_{h\in\cS} w_h\by^{(h)}}{\sum_{h\in\cS} w_h} - \bpi \right) \right\},
\nonumber
\end{equation}
where $\limvar$ denotes the asymptotic covariance matrix, and $h$ is the index for the units in the sample $\cS=\{ \by^{(h)}\}_{h=1}^n$.
Using a usual linearization argument for a ratio, we may write
\begin{equation}
\bSigma = \limvar \left\{ \sqrt{n} \,
\frac{ \sum_{h \in \cS} w_h ({\by}^{(h)}-\bpi)}{\E(\sum_{h \in \cS} w_h)}
\right\}.
\nonumber
\end{equation}
We consider a stratified multistage sampling scheme where the strata are labelled $a$ and the primary sampling units are labelled $b=1, \ldots, n_a$, where $n_a$ is the number of primary sampling units selected in stratum $a$. 

Then we write
\begin{equation}
\frac{\sum_{h\in\cS} w_h(\by^{(h)} - \bpi)}{\E (\sum_{h \in \cS} w_h)}
= \sum_{a=1} \sum_{b=1} \tilde{\bu}_{ab},
\nonumber
\end{equation}
where $\tilde{\bu}_{ab}=\sum_{h \in \cS} w_h({\by}^{(h)}-\bpi)/ 
\E(\sum_{h \in \cS} w_h)$ and $\cS$ is the set of sample units contained within primary sampling unit $b$ within stratum $a$.
So
\begin{equation}
\bSigma = \limvar \left\{ \sqrt{n}\sum_{a}\sum_{b} \tilde{\bu}_{ab} \right\}.
\nonumber
\end{equation}
A standard estimator of $n^{-1}\bSigma$ is the between-cluster variance estimator, which gives consistent variance estimates of linear and non-linear statistics when the number of clusters grows large (\cite[see e.g.][]{binder:83,bieler.williams:95,
williams:00,skinner:89, asparouhov2005sampling,wolter:07}) is then given by
\begin{equation}
n^{-1}\hat{\bSigma}=\sum_{a=1}^M\frac{n_a}{n_a-1}\sum_{b=1}^{n_a}(\textbf{u}_{ab}-\bar{\textbf{u}}_a)(\bu_{ab}-\bar{\bu}_a)',
\nonumber
\end{equation}
where $n_a$ is the number of sampled clusters from stratum $a$, $M$ is the total number of strata,  $\bu_{ab}=\sum_{h \in s} w_h({\by}^{(h)}-\bpi_0) / \sum_{h \in s} w_h$ is the total for all individuals $h$ in cluster $b$ in stratum $a$ and $\bar{\textbf{u}}_a = n_a^{-1} \sum_{b=1}^{n_a} \bu_{ab}$ is the average of $\bu_{ab}$ .
Here, $\bpi_0$ refers to the value of $\bpi$ under a simple null hypothesis or estimates of the model-implied proportions $\bpi(\btheta)$ under a composite null hypothesis.

In the case of simple random sampling, we have just one stratum, the clusters are elements and the weights are constant so that $n_a=n$, $\bu_{ab}=({\bf y}^{(h)}-\bpi_0)/n$, $\bar{\textbf{u}}_a=\bzero$ and the estimator reduces to $\frac{n}{n-1}\left(\diag(\bpi_0)-\bpi_0 \bpi_0'\right)$.

We only require  $\hat{\bSigma}_2=\bT_2  \hat{\bSigma} \bT_2'$ to compute the Wald and Pearson test statistics. We may write 
\begin{equation}
n^{-1} \hat{\bSigma}_2 =
\sum_{a=1}^M \frac{n_a}{n_a-1}
\sum_{b=1}^{n_a} (\textbf{v}_{ab}-\bar{\textbf{v}}_a)(\textbf{v}_{ab}-\bar{\textbf{v}}_a)'
\nonumber
\end{equation}
where $\textbf{v}_{ab}=\sum_{h \in \cS }w_h({\bf y}^{(h)}_2-\bpi_{20})/(\sum_{h \in \cS}w_h)$, $\bar{\textbf{v}}_a = n_a^{-1}\sum_{b=1}^{n_a} \textbf{v}_{ab}$, and
${\by}^{(h)}_2 = \bT_2{\by}^{(h)}$ is the $S \times 1$ indicator vector containing values $[y^{(h)}_i=1]$ and $[y^{(h)}_i=y^{(h)}_j=1]$ for different values of $i$ and $j$.

\section{Simulation study}
\label{simulation}
Our simulation study comprises three parts. 
The first part (Part A) focuses on assessing the performance of the parameter estimates and their asymptotic standard errors in bias and mean square error under an informative sampling scheme. 
These evaluations are carried out for the weighted pairwise likelihood, which accounts for unequal selection probabilities due to sampling design or informative sampling.
The simulation's second and third parts (Parts B and C) study the performance of the proposed test statistics about type I error and power for simple random sampling (SRS) and complex sampling, respectively.
 
We made a deliberate choice to include simulation results for SRS. 
This was done to evaluate the performance of the proposed test statistics under limited-information estimation methods. 
Research related to limited-information goodness-of-fit test statistics has primarily focused on full information maximum likelihood methods, as these produce the best asymptotically normal (BAN) estimators with the lowest asymptotic variance. 
In contrast, limited-information estimation methods do not yield BAN estimators.

Additionally, we investigate the degree to which the null distribution of the test statistics approximate the assumed theoretical distribution as we increase the sample size and the number of observed variables. 

\subsection{Simulation Part A: Assessing parameter estimates and standard errors for the weighted pairwise likelihood}

The data was generated as follows.
A population of size $N$ was created under the one-factor model using the specified true parameter values.
The size of $N$ varied according to the size of the sample $n$  to keep the sample-to-population ratio small and constant at $n/N = 1\%$ to avoid the need for any finite population correction factors.
Each individual indexed by $h$ was assigned a probability of selection denoted by $\pi_h$, which was set as $\pi_h = 1 / (1 + \exp(y^*_1))$ based on the first underlying variable $y^*_1$.
Thus, a larger value of $y^*_1$ resulted in a smaller probability of selection. 
A similar sampling scheme was described in \textcite{asparouhov2005sampling}.
The one-factor model was subsequently estimated using two different approaches, namely the unweighted pairwise maximum likelihood (PML) given in \eqref{eq:log_likel} and the weighted 
pairwise maximum likelihood (PMLW) given in \ref{eq:cxlog_likel}.
The sampling and estimation were replicated 1000 times and repeated for three sample sizes: $n=500, 1000, 5000$.

Table \ref{tbl-simA-bias} shows the parameter bias for different sample sizes. When the weights from the informative sampling are considered (PMLW), the bias is minimal and close to zero. Ignoring the weights (PML) leads to a more significant bias for the threshold estimates. 

\begin{longtable}{l|>{\raggedleft\arraybackslash}p{2cm}rrrrrr}

\caption{\label{tbl-simA-bias}Simulation A: Bias of the estimated factor loadings and thresholds for the one-factor model under an informative sampling scheme, $n=500,1000,5000$,  unweighted and weighted pairwise maximum likelihood (PML and PMLW).}

\tabularnewline

\toprule
\multicolumn{1}{l}{} &  & \multicolumn{2}{c}{\(n = 500\)} & \multicolumn{2}{c}{\(n = 1000\)} & \multicolumn{2}{c}{\(n = 5000\)} \\ 
\cmidrule(lr){3-4} \cmidrule(lr){5-6} \cmidrule(lr){7-8}
\multicolumn{1}{l}{} & True values & PML & PMLW & PML & PMLW & PML & PMLW \\ 
\midrule
\multicolumn{8}{l}{\textbf{Loadings}} \\ 
\midrule
\(\lambda_1\) & 0.80 & $-0.026$ & $-0.002$ & $-0.030$ & $-0.006$ & $-0.017$ & $0.005$ \\ 
\(\lambda_2\) & 0.70 & $-0.026$ & $-0.007$ & $-0.020$ & $-0.002$ & $-0.029$ & $-0.005$ \\ 
\(\lambda_3\) & 0.47 & $-0.020$ & $-0.001$ & $-0.024$ & $-0.004$ & $-0.024$ & $-0.003$ \\ 
\(\lambda_4\) & 0.38 & $-0.017$ & $-0.001$ & $-0.019$ & $-0.002$ & $-0.017$ & $0.003$ \\ 
\(\lambda_5\) & 0.34 & $-0.003$ & $0.010$ & $-0.019$ & $-0.002$ & $-0.022$ & $-0.006$ \\ 
\midrule
\multicolumn{8}{l}{\textbf{Thresholds}} \\ 
\midrule
\(\tau_1\) & $-1.43$ & $0.309$ & $-0.004$ & $0.311$ & $0.000$ & $0.304$ & $-0.007$ \\ 
\(\tau_2\) & $-0.55$ & $0.215$ & $-0.006$ & $0.218$ & $0.002$ & $0.213$ & $-0.008$ \\ 
\(\tau_3\) & $-0.13$ & $0.145$ & $-0.006$ & $0.161$ & $0.009$ & $0.153$ & $-0.001$ \\ 
\(\tau_4\) & $-0.72$ & $0.123$ & $0.003$ & $0.117$ & $-0.001$ & $0.120$ & $-0.002$ \\ 
\(\tau_5\) & $-1.13$ & $0.112$ & $0.009$ & $0.114$ & $0.009$ & $0.106$ & $0.002$ \\ 
\bottomrule

\end{longtable}
Table \ref{tbl-simA-se} gives the coverage and ratio of the standard deviation of the estimated parameters across replications to the estimated asymptotic standard error. The latter was calculated using the Godambe information matrix. The coverage is 0.95 for most parameters regardless of the sample size under the PMLW estimation.
Furthermore, the estimated asymptotic standard errors computed using the Godambe information matrix closely match the standard deviations computed across the 1000 replications.
Ignoring the weights during estimation impacts coverage and standard errors, leading to incorrect inferences.

\begin{sidewaystable}[ph!]

\begin{longtable}{l|rrrrrrrrrrrr}

\caption{\label{tbl-simA-se}Simulation A: Coverage rate and ratio of standard deviation across replications to the estimated asymptotic standard error (SD/SE) for factor loadings and thresholds for the one-factor model under an informative sampling scheme, $n=500,1000,5000$,  unweighted and weighted pairwise maximum likelihood (PML and PMLW).}

\tabularnewline

\toprule
\multicolumn{1}{l}{} & \multicolumn{4}{c}{\(n = 500\)} & \multicolumn{4}{c}{\(n = 1000\)} & \multicolumn{4}{c}{\(n = 5000\)} \\ 
\cmidrule(lr){2-5} \cmidrule(lr){6-9} \cmidrule(lr){10-13}
\multicolumn{1}{l}{} & \multicolumn{2}{c}{Coverage} & \multicolumn{2}{c}{SD/SE} & \multicolumn{2}{c}{Coverage} & \multicolumn{2}{c}{SD/SE} & \multicolumn{2}{c}{Coverage} & \multicolumn{2}{c}{SD/SE} \\ 
\cmidrule(lr){2-3} \cmidrule(lr){4-5} \cmidrule(lr){6-7} \cmidrule(lr){8-9} \cmidrule(lr){10-11} \cmidrule(lr){12-13}
\multicolumn{1}{l}{} & PML & PMLW & PML & PMLW & PML & PMLW & PML & PMLW & PML & PMLW & PML & PMLW \\ 
\midrule
\multicolumn{13}{l}{\textbf{Loadings}} \\ 
\midrule
\(\lambda_1\) & $0.95$ & $0.96$ & $1.01$ & $0.99$ & $0.94$ & $0.95$ & $1.04$ & $1.01$ & $0.89$ & $0.95$ & $1.22$ & $1.01$ \\ 
\(\lambda_2\) & $0.95$ & $0.95$ & $1.03$ & $0.99$ & $0.93$ & $0.95$ & $1.09$ & $1.01$ & $0.85$ & $0.95$ & $1.31$ & $0.99$ \\ 
\(\lambda_3\) & $0.95$ & $0.96$ & $1.02$ & $0.97$ & $0.94$ & $0.95$ & $1.06$ & $0.97$ & $0.85$ & $0.95$ & $1.35$ & $0.99$ \\ 
\(\lambda_4\) & $0.94$ & $0.94$ & $1.04$ & $1.03$ & $0.94$ & $0.95$ & $1.05$ & $1.00$ & $0.88$ & $0.95$ & $1.25$ & $1.01$ \\ 
\(\lambda_5\) & $0.95$ & $0.95$ & $0.99$ & $0.98$ & $0.95$ & $0.96$ & $1.02$ & $1.00$ & $0.90$ & $0.95$ & $1.19$ & $1.01$ \\ 
\midrule
\multicolumn{13}{l}{\textbf{Thresholds}} \\ 
\midrule
\(\tau_1\) & $0.01$ & $0.96$ & $4.38$ & $0.98$ & $0.00$ & $0.96$ & $6.24$ & $0.97$ & $0.00$ & $0.96$ & $13.81$ & $0.97$ \\ 
\(\tau_2\) & $0.04$ & $0.95$ & $3.91$ & $1.05$ & $0.00$ & $0.94$ & $5.47$ & $1.01$ & $0.00$ & $0.94$ & $12.03$ & $1.06$ \\ 
\(\tau_3\) & $0.22$ & $0.96$ & $2.93$ & $1.02$ & $0.03$ & $0.94$ & $4.03$ & $1.02$ & $0.00$ & $0.96$ & $8.76$ & $0.96$ \\ 
\(\tau_4\) & $0.49$ & $0.94$ & $2.22$ & $1.03$ & $0.20$ & $0.95$ & $2.97$ & $1.03$ & $0.00$ & $0.95$ & $6.40$ & $1.04$ \\ 
\(\tau_5\) & $0.61$ & $0.94$ & $1.88$ & $1.02$ & $0.42$ & $0.95$ & $2.39$ & $1.01$ & $0.00$ & $0.95$ & $5.13$ & $1.04$ \\ 
\bottomrule

\end{longtable}

\end{sidewaystable}

\subsection{Simulation Part B: Assessing performance of limited-information test statistics  under simple random sampling}
We evaluate the performance of the six test statistics outlined in Table \ref{tab:summaryteststats} under simple random sampling (SRS) and pairwise maximum likelihood estimation. We study five different models described in Table \ref{tab:models} with varying numbers of items ($p$) and factors ($q$) and five sample sizes: $n = 500, 1000, 2500, 5000$, and $10000$. This results in 25 simulation conditions. We conducted 1000 replications for each simulation condition.

\begin{table}[ht]
\centering
\caption{Simulations B and C: Models used for generating the data.}

\begin{tabular}{cll}
\toprule
Model & Label &Number of variables ($p$) and factors ($q$) \\ \midrule
1 & 1F 5V & $p=5$ and $q=1$ \\
2 & 1F 8V &$p=8$ and $q=1$ \\
3 & 1F 15V & $p=15$ and $q=1$ \\
4 & 2F 10V & $p=10$ and $q=2$, 5 indicators per factor \\
5 & 3F 15V & $p=15$ and $q=3$, 5 indicators per factor \\ \bottomrule
             \end{tabular}
    \label{tab:models}
\end{table}

In Model 1, the true factor loadings $\blambda'$ are $(0.8, 0.7, 0.47, 0.38, 0.34)$. The true threshold values $\btau'$ are $(-1.43, -0.55, -0.13, -0.72, -1.13)$.
For Model 2, the factor loading values are identical to those in  Model 1 for the first five items. Additionally, the first three factor loadings (0.8, 0.7, 0.47) from Model 1 are repeated to set the values for the last three items in Model 2. The threshold values follow the same pattern. The same mechanism is used to set the true values for Model 3.
Models 4 and 5 are confirmatory factor analysis models. The true factor loadings for each factor are used in Model 1. The same goes for the thresholds.
For Models 4 (two factors) and 5 (three factors), the factor correlations are $\rho = 0.3$ and $\brho' = (0.2, 0.3, 0.4)$, respectively.

In each replication within each condition, we compute the goodness-of-fit test statistic using equation \eqref{eq:quadformtest}, and the weight matrices used are summarized in Table \ref{tab:summaryteststats}.
The moment matching adjustment described in Appendix \ref{sec:estchisq} has been applied to all tests but the Wald and Wald (VCF). The degrees of freedom for the Wald test statistics are in the range $(S-m, S)$. We found that $S-m$ gave the best results in approximating the test statistic distribution, which we used in all our simulations. Our comparison focuses on the Wald, WaldVCF and the Pearson test statistics. The study also includes the performance of simpler versions of the tests (WaldDiag, RSS and Multinomial). However, we recommend using these simpler versions independently only after further investigations to determine their suitability in any model setting. 

\begin{table}[htb]
\centering
\caption{Summary of the test statistics evaluated in simulations B and C.}
\label{tab:summaryteststats}
\begin{tabular}{@{}lllll@{}}
\toprule
  & Name                & Weight Matrix ($\bXi$)        & d.f.                   \\ \midrule
1 & Wald                & $\bOmega_2^+$                 & $S-m, S$   \\
2 & Wald (VCF)          & $\bXi\bOmega_2\bXi$           & $S-m$                        \\
3 & Wald Diagonal     & $\diag(\bOmega_2)^{-1}$       & est.    \\
4 & Pearson             & $\diag(\bpi_2)^{-1}$ & est.    \\
5 & RSS             & $\bI$ & est.    \\
6 & Multinomial             & $[\diag(\bpi_2) - \bpi_2\bpi_2']^{-1}$ & est.   \\
\bottomrule
\end{tabular}
\end{table}

To conduct a power analysis, we introduced an additional latent variable $z \sim \operatorname{N}(0,1)$ independent of the latent variables $\bfeta$. This extra variable was included in the data generating model to capture together with $\bfeta$ the associations among the $y^*$ variables.
The loadings of variable $z$ in the data-generating model closely resembled the loadings of the true latent factor $
\bfeta$. 
For Models 1 and 2, the extra latent variable $z$ loads onto all items except two ($y_2$ and $y_6$).
For Model 3, $z$ loads onto all items except three ($y_2, y_6, y_{14}$).
For Models 4 and 5, $z$ loads onto all items.
This additional latent variable in the data generation process results in a misspecified model, and the power of the test correctly detects the misspecification.

\begin{figure}[htbp]
\centering
\includegraphics[width=0.96\textwidth]{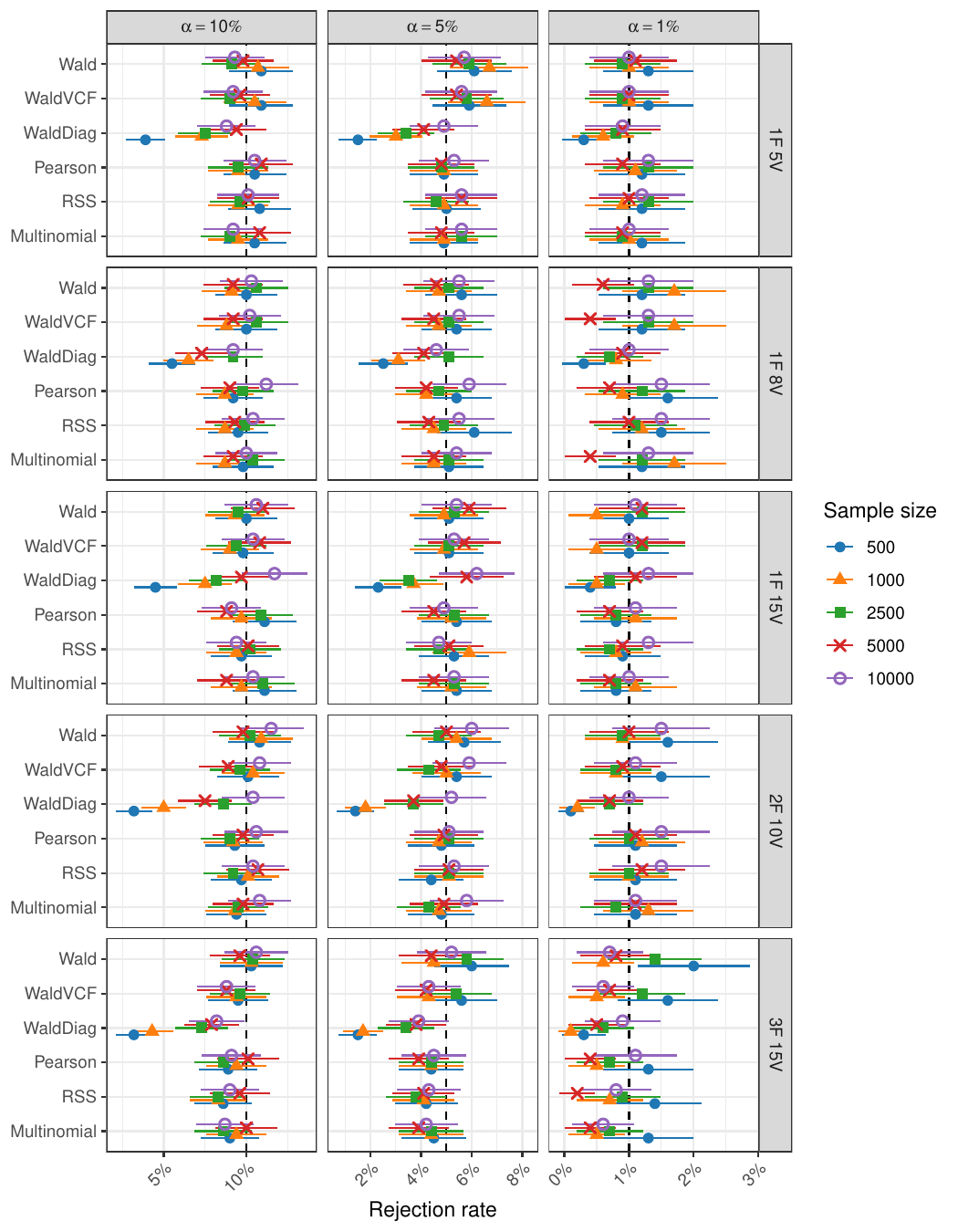}
\caption{Simulation B: Type I error rates at significance levels $\alpha=(10\%, 5\%, 1\%)$ for the six test statistics in Table \ref{tab:summaryteststats} across the five models in Table \ref{tab:models} and $n=500$, $1000$, $2500$, $5000$, $10000$, SRS. The reported intervals are the 95\% confidence intervals for each rejection proportion.}
\label{fig:resultsrstype1}
\end{figure}

\begin{figure}[htbp]
\centering
\includegraphics[width=0.96\textwidth]{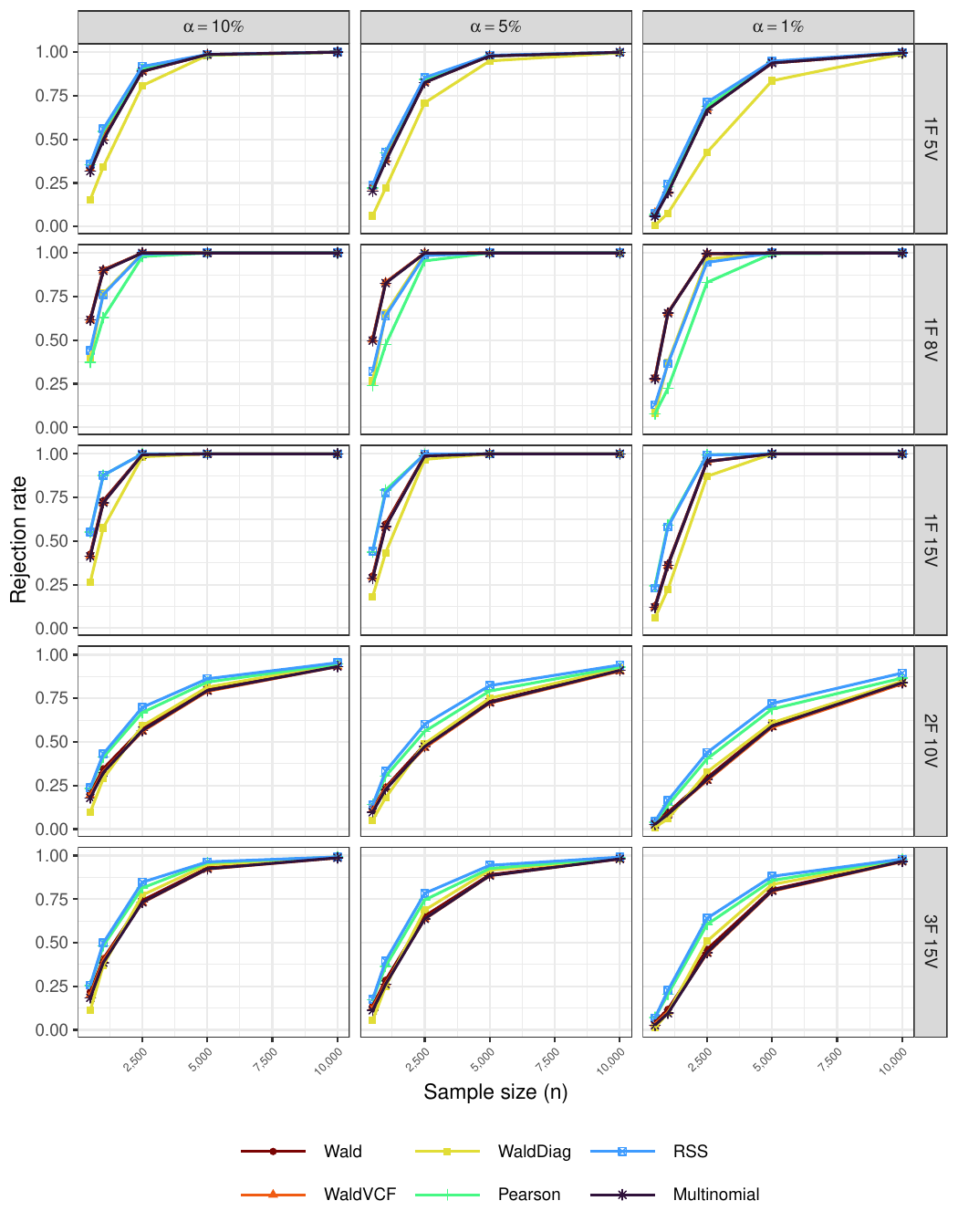}
\caption{Simulation B: Power analysis at significance levels $\alpha=(10\%, 5\%, 1\%)$ for the six test statistics in Table \ref{tab:summaryteststats} across the five models in Table \ref{tab:models}, and $n=500$, $1000$, $2500$, $5000$, $10000$, SRS. }
\label{fig:resultsrspower}
\end{figure}

Figure \ref{fig:resultsrstype1} gives the type I error rates for the six test statistics across all simulation conditions.  The Wald Diagonal test statistic exhibited the poorest performance. In contrast, Pearson and the Wald-type test statistics demonstrated satisfactory performance across all three significance levels $\alpha=10\%, 5\%$, and $1\%$, with improvement noted as the sample size increased.  The power of all tests, as shown in Figure \ref{fig:resultsrspower}, increases with sample size, although it remained at lower levels in the case of the two- and three-factor models.
In addition, as the complexity of the model increased, the power of the Wald-type test statistics consistently decreased compared to the Pearson test statistic. This could be attributed to the instability in inverting the $\hat{\Omega}_2$ matrix and potential issues with its estimation arising from possible sparsity in the lower order margins and smaller sample sizes.

\subsection{Simulation C: Assessing performance of limited-information test statistics  under complex sampling}

We generated data for an entire population, mirroring a commonly employed sampling approach in educational surveys. 
The population comprises 2,000 schools, acting as Primary Sampling Units (PSUs). 
These schools are stratified into three types: ``A'' (400 units), ``B'' (1,000 units), and ``C'' (600 units). 

The schools are of different sizes so that an unequal probability sample may be conducted for the clustered sample scenario based on each school's size.
The number of students allocated to each school follows a normal distribution, with a mean of 500 and a standard deviation of 125. 
Each school's assigned students is then rounded to the nearest whole number.

Additionally, students within each school type are randomly distributed among classes, where the average class sizes are set at 15, 25, and 20 for school types A, B, and C, respectively.
The simulated population comprised approximately one million students (clustered within classrooms and classrooms clustered within schools).
The population statistics, encompassing the count of schools, school types, and average class sizes, are given in Table \ref{tab:summarysimpop}.

\begin{table}[ht]
\centering
\caption{Simulation C: Population statistics for the simulated school population.}
\label{tab:summarysimpop}
\begin{tabular}{crccrrrrrrrrrr}
\toprule
\multicolumn{2}{c}{Schools} && \multicolumn{5}{c}{Students} && \multicolumn{4}{c}{Class size} \\
\cline{1-2} \cline{4-8} \cline{10-13}
Type & $N$ && $N$ & Avg. & SD & Min. & Max. && Avg. & SD & Min. & Max. \\
\midrule
A & 400 && 200,670 & 501.7 & 96.9 & 253 & 824 && 15.2 & 3.9 & 3 & 36 \\
B & 1000 && 501,326 & 501.3 & 100.2 & 219 & 839 && 25.7 & 5.0 & 6 & 48 \\
C & 600 && 303,861 & 506.4 & 101.9 & 195 & 829 && 20.4 & 4.4 & 6 & 39 \\
\bottomrule
\end{tabular}
\end{table}
The item responses are generated based on the five factor models outlined in Table \ref{tab:models}. 
The PSU (schools) are stratified into distinct strata based on the latent vector $\bfeta$,  representing the abilities of each student.
The observation units are ordered in descending order based on their latent variable value for the first factor $\eta_{1h}$.
Subsequently, they are allocated to school types A, B, or C. 
If the vector of observed variables ${\bf y}$ represents test items, this categorisation implies that school type A comprises `high' ability students, type B `medium' ability students, and type C `low' ability students. 
This data-generating scheme results in varying intraclass correlations (ICC) for response items across PSUs, ranging from 0.05 to 0.6.
Of note, \textcite{hedges2007intraclass} found that a typical value for the ICC in an educational study was about 0.16, adjusted for covariates.

To obtain a sample from this simulated population, we explore two multistage probability sampling designs: 1) two-stage cluster sampling and 2) stratified cluster sampling.
The following discussion details the process of selecting a sample of (approximate) size $n$ from each design.
Table \ref{tab-npsu} shows the number of clusters (PSUs) sampled using each sampling design discussed below.

\begin{enumerate}

\item \textbf{Two-stage cluster sampling}

Let \( n_c = \lfloor \frac{n}{21.5} \rfloor \), where 21.5 represents the average class size across all types of schools in the population. 
The value of \( n_c \) approximates the number of schools (PSUs). Schools are selected using Probability Proportional to Size (PPS) sampling, where the size variable is the number of students in each school.
Next, one class is selected using simple random sampling from each selected PSU.
All students in the selected class are included in the sample.
The probability of selecting a student from school $b$ is:
\[
\Pr(\text{selection}) = \frac{z_b}{\sum_{b} z_b} \times \frac{1}{M_b},
\]
where $z_b$ is the number of students in school $b$, $\sum_{b} z_b$ is the total number of students in all schools in the population, and $M_b$ is the number of classes in school $b$.

The total sample size varies from sample to sample due to varying school and class sizes.
However, on average, it is expected to be close to \( n = n_c \times 21.5 \), where 21.5 is the average number of students per class.

\item \textbf{Stratified cluster sampling}

Select a fixed number of schools (PSU) denoted by $n_c$ using SRS from each stratum.
The quantity $n_c$ is determined using the formula
$n_c = \lfloor n / (15 + 20 + 25) \rfloor$, since we want the sample size $n$ to be close to $15n_c + 20n_c + 25n_c$, where the numbers 15, 20 and 25 represent the average class size in each stratum.
One class is selected from each selected school using SRS. All the students from the selected class are included in the sample. 

The inclusion probability for a student from school, $b$ within stratum $a$, is calculated to be:
$$
\Pr(\text{selection}) = \frac{n_a}{N_a} \times \frac{n_{ab}}{N_{ab}} \times \frac{n_{abc}}{N_{abc}},
$$
where $n_a$ ($N_a$) is the sample (population) number of schools in stratum $a$, 
$n_{ab}$ ($N_{ab}$) is the sample (population) number of classes in school $b$ within stratum $a$, and 
$n_{abc}$ ($N_{abc}$) is the sample (population) number of students in class $c$, school $b$ within stratum $a$. In our experiment, $n_a$ was chosen to be the same in each stratum (equal allocation) denoted with $n_c$, and all students were selected from the selected classes, $n_{abc}=N_{abc}$.

\end{enumerate}

\begin{table}[htb]
\caption{Number of clusters (PSUs) sampled with each sampling method.}
\label{tab-npsu}
\centering
\begin{tabular}[t]{lrrrrr}
\toprule
Sampling method & $n = 500$ & $n = 1000$ & $n = 2500$ & $n = 5000$ & $n = 10000$\\
\midrule
Cluster & 23 & 47 & 116 & 233 & 465\\
Stratified-cluster & 24 & 51 & 126 & 249 & 501\\
\bottomrule
\end{tabular}
\end{table}

For the two-stage cluster sampling design, we provide additional results in Appendix C. 
Table \ref{tbl-clust-se} shows the coverage and the SD/SE ratio for each estimated parameter obtained from 1000 replications over the estimated asymptotic standard error for all model parameters. 
This is for the one-factor model with five items, using weighted pairwise likelihood and considering clustering in estimating the variability matrix $\bJ$ \eqref{eq:clusterJ} in the Godambe information matrix. 
The results show that when clustering is considered in estimating the Godambe information matrix, the standard deviation to standard error ratio is much closer to one than when it is not accounted for.

\begin{figure}
\centering
\includegraphics[width=0.94\textwidth]{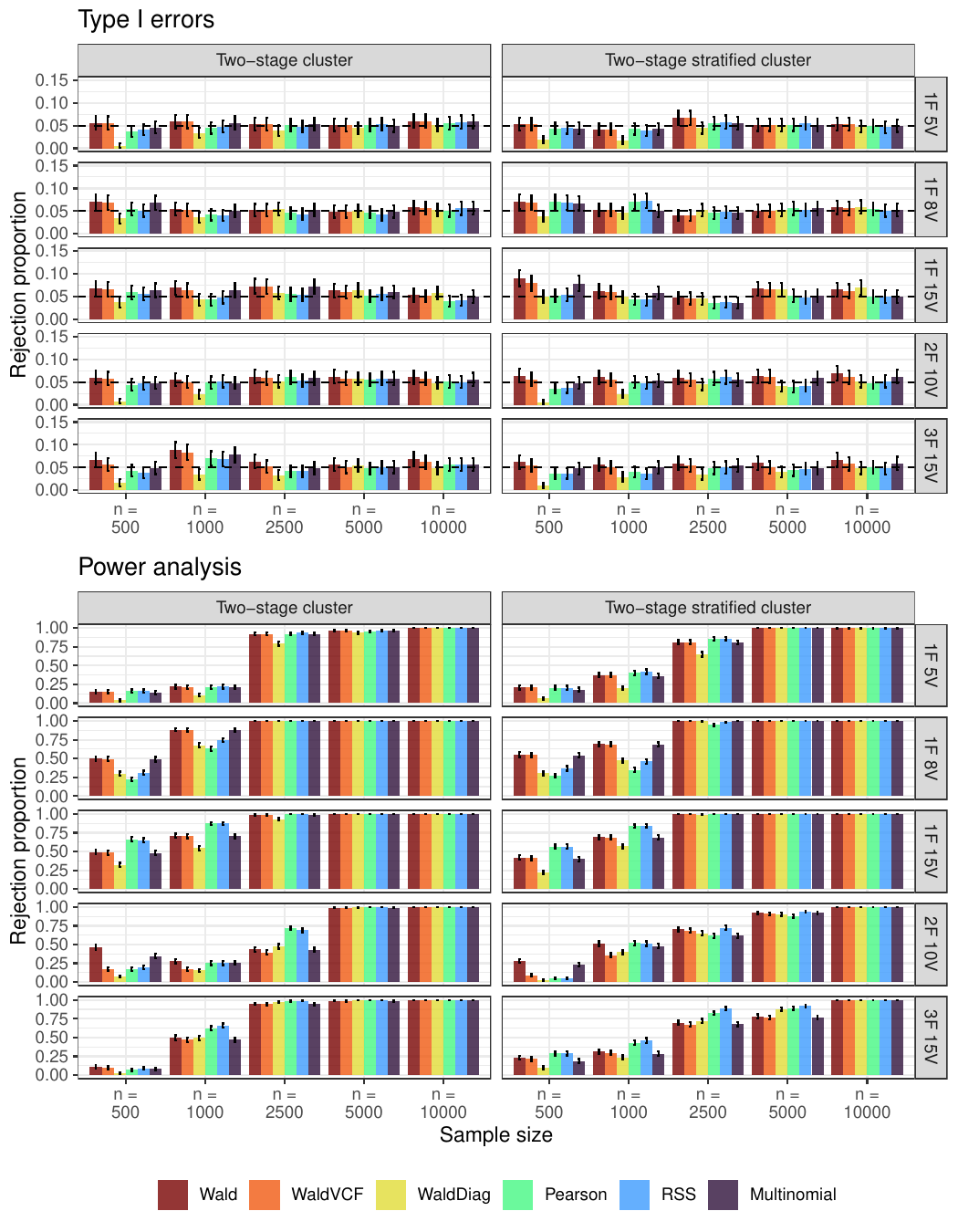}
\caption{Simulation C: Type I error rates (top) and power analysis (bottom) for the two complex sampling designs at significance level $\alpha=5\%$ for the six test statistics in Table \ref{tab:summaryteststats} across the five models in Table \ref{tab:models}, and $n=500,1000,2500,5000,10000$. The reported intervals are the 95\% confidence intervals for each rejection proportion.}
\label{fig:resultcomplextype1}
\end{figure}

Figure \ref{fig:resultcomplextype1} displays the type I error and power rates for the six test statistics under the two complex sampling designs across the five models outlined in Table \ref{tab:models}.
All test statistics performed well under both complex designs, except for the WaldDiag, which required larger sample sizes. Due to its poor performance even under SRS, it is not advisable to use the WaldDiag. The Wald and WaldVCF also exhibited poor performance in the larger models with smaller sample sizes. The Wald test is known to be unstable, mainly when used with data from complex sampling designs and has poor small sample behaviour (\cite[see e.g.][]{fay:85,lumley.scott:14}). As  \textcite{fay:85} and \textcite{skinner:19} have stated, the covariance matrix estimator of the estimated proportions is consistent, but in practice, it will typically be less precise than the multinomial analogues. This reduced precision seriously affects the inversion required in the Wald statistic. In particular, \textcite[][p. 148]{fay:85} wrote that `the instability in the estimated inverse, in turn, inflates the rate of rejection under the null hypothesis, often enough to make the test unusable'. 

Similar to SRS, the power of the test statistics increased with sample size across all models. Notably, the 2F10V model required a sample size 5000 to exhibit larger power levels in both sampling designs. The 3F15V model also needed a large sample size under the two-stage stratified cluster design. For sample sizes 500 and 1000, the test statistics exhibited low power. It is important to note that in those cases, the Pearson test statistic exhibited higher or similar power than the Wald-type tests except for the 1F8V model under both designs. The Wald-type test statistics are possibly affected by the instability of computing and inverting the $\hat{\bOmega}_2$ matrix and the likely sparsity even in the lower order margins in the bigger models and small sample sizes.

The asymptotic distribution of the proposed test statistics is checked using QQ plots. In Appendix \ref{apx:hist}, Figures \ref{fig:histteststat1}, \ref{fig:histteststat2} and \ref{fig:histteststat3} display QQ plots across the 1000 replications for all five models, five sample sizes, and simple random sampling, two-stage cluster sampling, and two-stage stratified cluster sampling respectively.
The plots reveal a generally close alignment between the theoretical distribution of each test and its Monte Carlo distribution. The largest deviations from the asymptotic distribution were found in the WaldDiag test.

\section{Discussion}
The paper expands the pairwise likelihood estimation and limited information test statistics to accommodate complex and informative sampling designs. Specifically,  sampling and informative weights are incorporated into the pairwise likelihood to consider the complex design during model estimation and the estimation of asymptotic standard errors. Additionally, the paper introduces limited-information test statistics of Wald and Pearson's types tailored for complex sampling within the pairwise likelihood estimation framework.

The weighted pairwise likelihood produced unbiased estimates and robustly estimated standard errors. The limited-information test statistics implemented using pairwise likelihood estimates showed a good performance in terms of type I error and power under simple random and two complex sampling designs, namely, two-stage cluster and stratified cluster sampling. 
Although we studied both limited information Wald and Pearson type test statistics that work satisfactorily in most cases, the Wald test statistic, for the reason we discussed above, can be unstable, mainly when used to data from complex sampling designs and has poor small sample 
behaviour. Therefore, we recommend using the Pearson test statistic with the pairwise likelihood estimator and the moment matching adjustment. Also, the Pearson test statistic does not require to invert the $\bOmega_2$ matrix. 

One must also consider that sparseness may occur in the lower-order margins for complex models and small sample sizes. In this case, further investigation is needed to study their performance.  

Further research can extend the tests to ordinal responses and investigate resampling methods for estimating the proportions' variance under multi-stage designs. A Bayesian pairwise approach can also be helpful here for obtaining the test statistics distribution via posterior sampling without relying on the large sample theory presented in this paper.

\section*{Supplementary Material}

All analyses were conducted in R \parencite{rsoftware} using the lavaan.bingof \parencite{jamil2024lavaanbingof} package. The R scripts are available at \url{https://osf.io/2d97y/}, including tabulated Type I error rates and power for all simulations conducted.

\printbibliography

\appendix

\section{Proof}
\label{apx:proof}

From \eqref{eqder2} we wrote that the score vector can be expressed as a transformation of the $(\bp_2 - \bpi_2(\btheta))$ vector, which we prove here.
Extending \eqref{eq:scorease2} to vector form, we have
\begin{align}
\bbR^{m} \ni \nabla\pl(\btheta)
&= n\left( \frac{\partial \pimodpl}{\partial\theta_k}  \right)_{\substack{i<j \\ k=1,\dots,m}}' \diag\big(\pimodpl\big)^{-1} \big( p_{c_ic_j}^{(ij)} -  \pimodpl \big)_{i<j} \nonumber \\
&= n\left( \frac{\partial \pimodpl}{\partial\theta_k}  \right)_{\substack{i<j \\ k=1,\dots,m}}' \diag\big(\pimodpl\big)^{-1} \, (\bBij)_{i<j} \big(\bp_2 - \bpi_2(\btheta)\big), \label{eq:vectorformscore}
\end{align}
where $(\bBij)_{i<j}$ is the $4p(p-1)/2 \times S$ matrix containing stacked $1\times S$ row vectors $\bBij$ satisfying \eqref{eqder1}.
That is, for any $i,j\in\{1,\dots,p\}$ and $i<j$,
$$
\begin{pmatrix}
\bB_{00}^{(ij)} \\
\bB_{10}^{(ij)} \\
\bB_{01}^{(ij)} \\
\bB_{11}^{(ij)} \\
\end{pmatrix}
\big(\bp_2 - \bpi_2(\btheta)\big)
=
\begin{pmatrix}
p_{00}^{(ij)} - \pi_{00}^{(ij)}(\btheta) \\
p_{10}^{(ij)} - \pi_{10}^{(ij)}(\btheta) \\
p_{01}^{(ij)} - \pi_{01}^{(ij)}(\btheta) \\
p_{11}^{(ij)} - \pi_{11}^{(ij)}(\btheta)
\end{pmatrix}.
$$
The design should be obvious and simple for the patterns $c_{i}c_{j}=10,01,11$.
As for the pattern `00', we may use the fact that the sum of the pairwise proportions is 1:
\begin{align*}
p_{00}^{(ij)} - \pi_{00}^{(ij)}(\btheta)
&= (1 - p_{10}^{(ij)} - p_{01}^{(ij)} - p_{11}^{(ij)}) - \big(1 - \pi_{10}^{(ij)}(\btheta)  - \pi_{01}^{(ij)}(\btheta) - \pi_{11}^{(ij)}(\btheta)\big) \\
&=
- \big(p_{10}^{(ij)} - \pi_{10}^{(ij)}(\btheta)\big)
- \big(p_{01}^{(ij)} - \pi_{01}^{(ij)}(\btheta)\big)
- \big(p_{11}^{(ij)} - \pi_{11}^{(ij)}(\btheta)\big) \\
&=
-\bB_{10}^{(ij)} \big(\bp_2 - \bpi_2(\btheta)\big)
-\bB_{01}^{(ij)} \big(\bp_2 - \bpi_2(\btheta)\big)
-\bB_{11}^{(ij)} \big(\bp_2 - \bpi_2(\btheta)\big) \\
&= \underbrace{- (\bB_{10}^{(ij)} + \bB_{01}^{(ij)} + \bB_{11}^{(ij)})}_{\bB_{00}^{(ij)}}\big(\bp_2 - \bpi_2(\btheta)\big).
\end{align*}
Combining this design matrix with the preceding terms in \eqref{eq:vectorformscore}, we get
\begin{align}
\bB(\btheta) := \left( \frac{\partial \pimodpl}{\partial\theta_k}  \right)_{\substack{i<j \\ k=1,\dots,m}}' \diag\big(\pimodpl\big)^{-1} \, (\bBij)_{i<j},
\end{align}
an $m \times S$ matrix dependent on the parameter vector $\btheta$.
The equation \eqref{eqder2} then follows directly.

\section{Estimation of degrees of freedom}
\label{sec:estchisq}

We discuss the moment matching procedure used to estimate the degrees of freedom of $X^2 = n\hat{\mathbf e}_2' \hat{\boldsymbol\Xi}\hat{\mathbf e}_2$. This statistic has a limiting distribution of $\sum_{s=1}^S \delta_s U_s$, where each $U_s \iid \chi^2_1$ and $\delta_s$ represents the eigenvalues of $\bM = \bOmega_2^{1/2}\bXi\bOmega_2^{1/2}$.

The eigenvalues, which are invariant under cyclic permutations, can be more efficiently computed as the eigenvalues of $\bXi\bOmega_2$, especially when $\bXi$ is diagonal and $\bOmega_2$ is dense.
In our context, these eigenvalues are estimated using the hat versions of $\bOmega_2$ and $\bXi$.
For the Wald and Wald VCF test, $\bM$ becomes idempotent, resulting in an exact chi-square distribution for $X^2$. Consequently, these estimation procedures are specifically applied to the diagonal Wald and the Pearson test statistics.

Suppose $Y$ follows a $\chi^2_c$ distribution. We assume \(X^2 \) can be approximately represented as a linear transformation of this chi-square random variate:
\begin{equation}
X^2 \approx a + bY.
\end{equation}
The first three moments of $X^2$ are expressed as \parencite[Theorem 3.2b.2, p. 53]{mathai1992quadratic}:
\begin{equation}
\mu_1(X^2) = \operatorname{tr}({\boldsymbol\Xi}{\boldsymbol\Omega}_2), \hspace{1em}
\mu_2(X^2) = 2\operatorname{tr}\big(({\boldsymbol\Xi}{\boldsymbol\Omega}_2)^2\big), \hspace{1em}
\mu_3(X^2) = 8\operatorname{tr}\big(({\boldsymbol\Xi}{\boldsymbol\Omega}_2)^3\big), \label{eq:moment1}
\end{equation}
The moments of the approximating random variable $a + bY$ are denoted as:
\begin{equation}
\mu_1(a+bY) = a + bc, \hspace{2em}
\mu_2(a+bY) = 2b^2c, \hspace{2em}
\mu_3(a+bY) = 8b^3c. \label{eq:moment2}
\end{equation}
The formulas for the raw moments of a chi-square random variable $Y$ with \(c\) degrees of freedom are
\(\E(Y^k)=c(c+2)(c+4)\cdots(c+2k-2)\).
By equating these moment expressions and solving for the parameters $a,b$ and $c$ in the \textbf{three-moment adjustment} equation:
\begin{equation}
b = \frac{\mu_3(X^2)}{4\mu_2(X^2)}, \hspace{2em}
c = \frac{\mu_2(X^2)}{2b^2}, \hspace{2em}
a = \E(X^2) - bc.
\nonumber
\end{equation}

We can also use two-moment and one-moment matching techniques similarly.
For the two-moment adjustment, if we assume \(X^2 \approx bY\) where \(Y\sim\chi^2_c\) (i.e. set $a=0$), the formulae for $b$ and $c$ are as follows.
\begin{equation}
b = \frac{\mu_2(X^2)}{2\mu_1(X^2)}
\hspace{1em}\text{and}\hspace{1em}
c = \frac{\mu_1(X^2)}{b}.
\nonumber
\end{equation}
On the other hand, in the \textbf{one-moment adjustment}, let's set $c=S$, which represents the number of degrees of freedom available for testing. In this case, $b$ can be calculated as: $b = \mu_1(X^2) / c$.

In these cases, the tail probabilities can be approximated as:
\begin{equation}
\Pr(X^2 > x) \approx \Pr\left(Y > \frac{x-a}{b} \ \Big| \ Y \sim \chi^2_c\right).
\nonumber
\end{equation}
The moment matching procedure is often referred to as the Satterwaithe approximation. The parameter estimates for $a,b,c$ are calculated using estimated versions of $\bXi$ and $\bOmega_2$ in the given formulas.
In this paper, we use the three-moment procedure due to previous findings suggesting inaccuracies in  $p$-values obtained from the one-moment procedure (e.g. \cite{maydeu2008overview}).

\begin{sidewaystable}[ph!]
\section{Two-stage cluster sampling design, coverage and standard errors for parameter estimates}
\label{apx:clusterse}

\begin{longtable}{l|rp{1.2cm}rp{1.2cm}rp{1.2cm}rp{1.2cm}rp{1.2cm}rp{1.2cm}}

\caption{\label{tbl-clust-se}Two-stage cluster sampling: Coverage rate and the ratio of standard deviation across replications to the estimated asymptotic standard error (SD/SE) for factor loadings and thresholds for the one-factor model, $n=500,1000,5000$, weighted and weighted adjusted for clustering pairwise maximum likelihood (PMLW and PMLW-adj.}

\tabularnewline

\toprule
\multicolumn{1}{l}{} & \multicolumn{2}{c}{Coverage} & \multicolumn{2}{c}{SD/SE} & \multicolumn{2}{c}{Coverage} & \multicolumn{2}{c}{SD/SE} & \multicolumn{2}{c}{Coverage} & \multicolumn{2}{c}{SD/SE} \\ 
\cmidrule(lr){2-3} \cmidrule(lr){4-5} \cmidrule(lr){6-7} \cmidrule(lr){8-9} \cmidrule(lr){10-11} \cmidrule(lr){12-13}
\multicolumn{1}{l}{} & \multicolumn{4}{c}{\(n = 500\)} & \multicolumn{4}{c}{\(n = 1000\)} & \multicolumn{4}{c}{\(n = 5000\)} \\ 
\cmidrule(lr){2-5} \cmidrule(lr){6-9} \cmidrule(lr){10-13}
\multicolumn{1}{l}{} & PMLW & PMLW-adj & PMLW & PMLW-adj & PMLW & PMLW-adj & PMLW & PMLW-adj & PMLW & PMLW-adj & PMLW & PMLW-adj \\ 
\midrule\addlinespace[2.5pt]
\multicolumn{13}{l}{\textbf{Loadings}} \\ 
\midrule\addlinespace[2.5pt]
\(\lambda_1\) & $0.94$ & $0.92$ & $1.08$ & $1.10$ & $0.93$ & $0.92$ & $1.09$ & $1.08$ & $0.94$ & $0.96$ & $1.04$ & $1.00$ \\ 
\(\lambda_2\) & $0.94$ & $0.93$ & $1.07$ & $1.04$ & $0.93$ & $0.94$ & $1.11$ & $1.05$ & $0.93$ & $0.95$ & $1.06$ & $0.99$ \\ 
\(\lambda_3\) & $0.93$ & $0.93$ & $1.11$ & $1.06$ & $0.92$ & $0.93$ & $1.14$ & $1.06$ & $0.92$ & $0.95$ & $1.08$ & $0.99$ \\ 
\(\lambda_4\) & $0.95$ & $0.93$ & $1.03$ & $1.03$ & $0.94$ & $0.94$ & $1.07$ & $1.03$ & $0.92$ & $0.94$ & $1.08$ & $1.02$ \\ 
\(\lambda_5\) & $0.92$ & $0.90$ & $1.08$ & $1.10$ & $0.93$ & $0.92$ & $1.08$ & $1.07$ & $0.95$ & $0.95$ & $1.02$ & $1.00$ \\ 
\midrule\addlinespace[2.5pt]
\multicolumn{13}{l}{\textbf{Thresholds}} \\ 
\midrule\addlinespace[2.5pt]
\(\tau_1\) & $0.64$ & $0.93$ & $2.19$ & $1.08$ & $0.67$ & $0.95$ & $2.09$ & $1.02$ & $0.66$ & $0.96$ & $1.99$ & $0.97$ \\ 
\(\tau_2\) & $0.56$ & $0.93$ & $2.63$ & $1.05$ & $0.57$ & $0.95$ & $2.45$ & $0.97$ & $0.57$ & $0.97$ & $2.41$ & $0.95$ \\ 
\(\tau_3\) & $0.70$ & $0.94$ & $1.85$ & $1.01$ & $0.71$ & $0.95$ & $1.86$ & $1.01$ & $0.71$ & $0.96$ & $1.84$ & $0.99$ \\ 
\(\tau_4\) & $0.80$ & $0.94$ & $1.57$ & $1.04$ & $0.81$ & $0.95$ & $1.45$ & $0.96$ & $0.80$ & $0.96$ & $1.51$ & $0.99$ \\ 
\(\tau_5\) & $0.84$ & $0.92$ & $1.38$ & $1.05$ & $0.87$ & $0.96$ & $1.28$ & $0.97$ & $0.87$ & $0.96$ & $1.31$ & $0.98$ \\ 
\bottomrule

\end{longtable}
\end{sidewaystable}

\section{QQ plots of test statistics under $H_0$}
\label{apx:hist}

\subsection*{Simple random sampling}
\begin{figure}[htbp]
\centering
\includegraphics[width=0.95\textwidth]{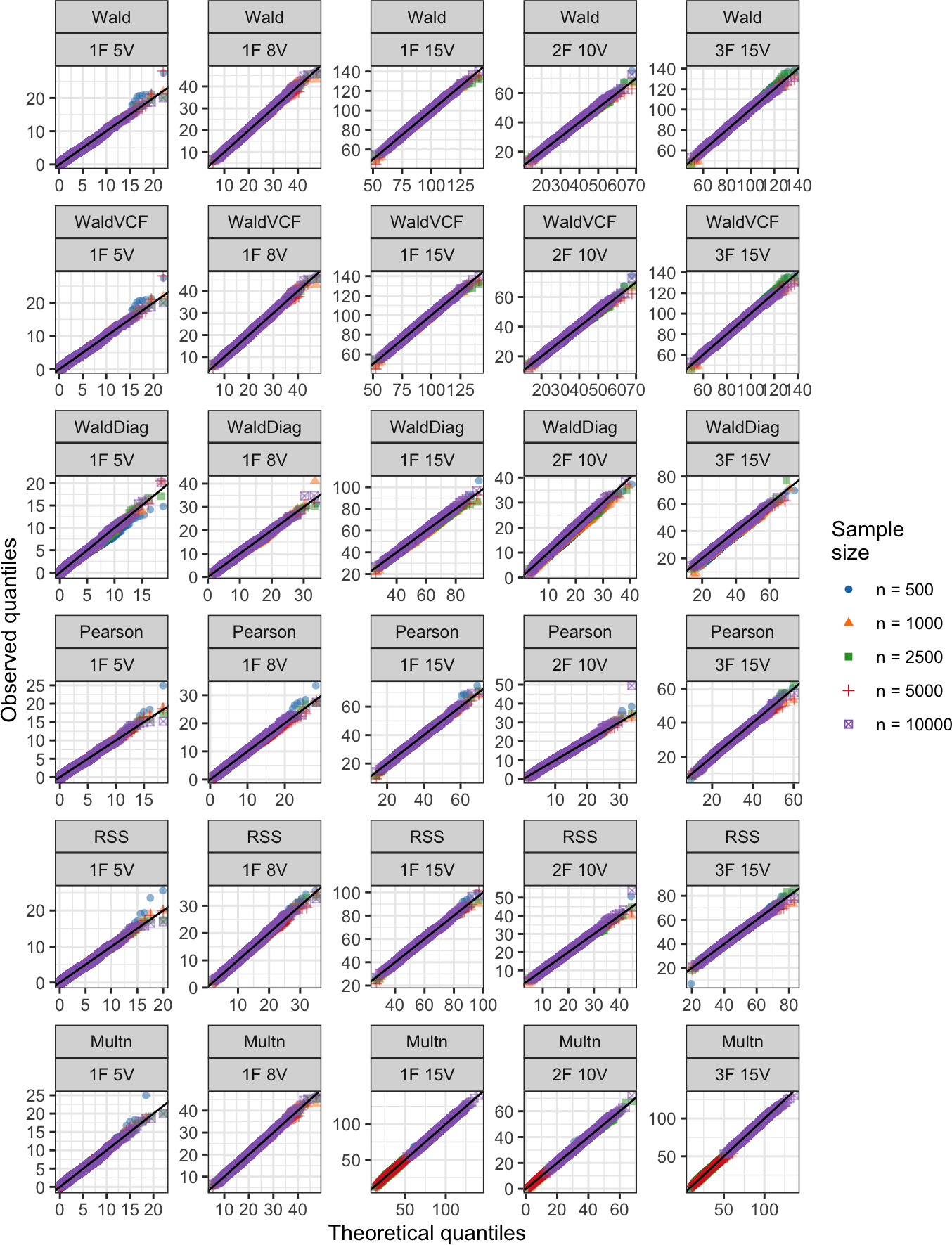}
\caption{QQ plot for the six test statistics in Table \ref{tab:summaryteststats} across the five models in Table \ref{tab:models} from simulated data for simple random sampling under the null hypothesis and their respective theoretical asymptotic distributions, $n=500,1000,2500,5000,10000$.}
\label{fig:histteststat1}
\end{figure}
\newpage
\subsection*{Two-stage cluster sampling}

\begin{figure}[htbp]
\centering
\includegraphics[width=1\textwidth]{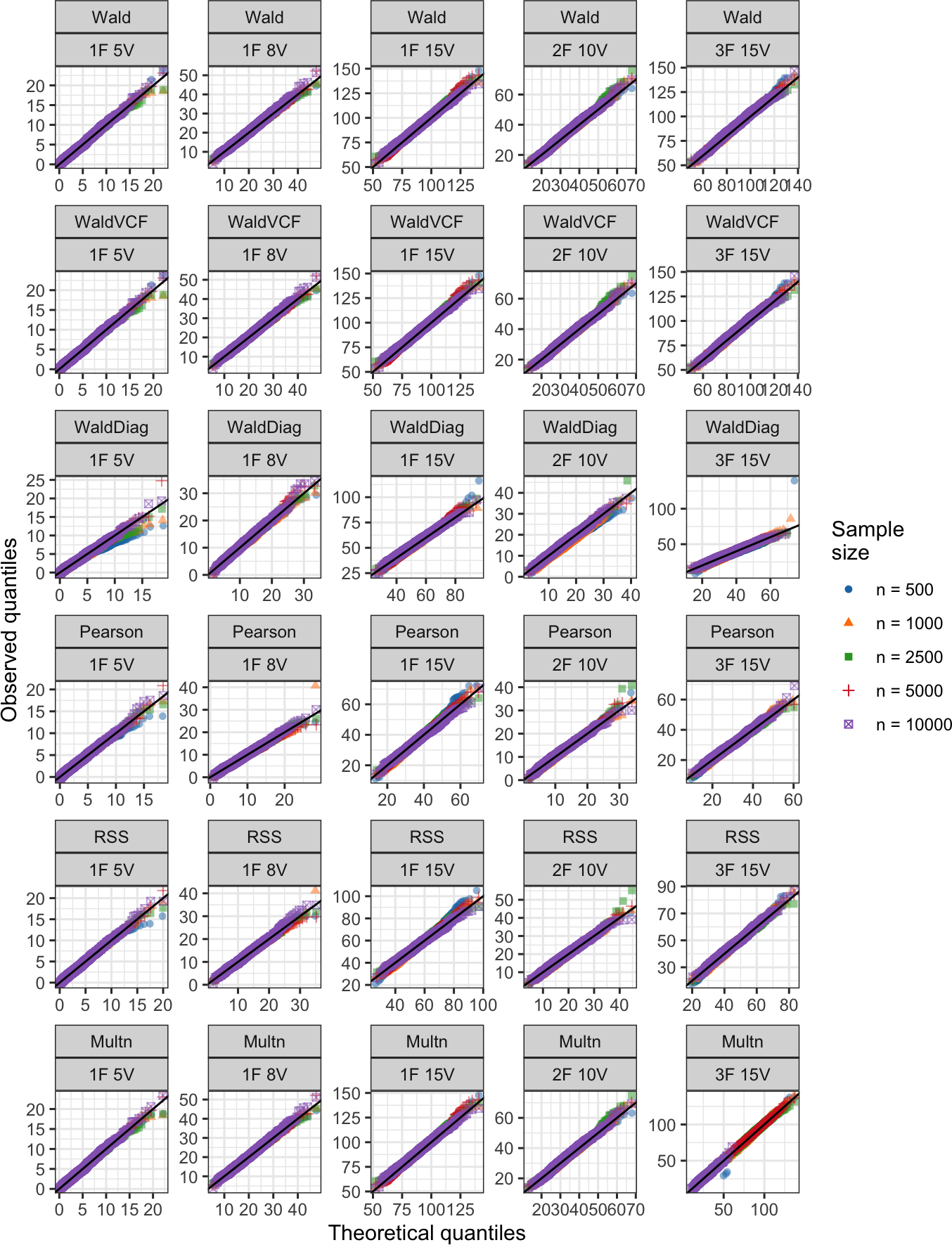}
\caption{QQ plot for the six test statistics in Table \ref{tab:summaryteststats} across the five models in Table \ref{tab:models} from simulated data for the two-stage cluster sampling under the null hypothesis and their respective theoretical asymptotic distributions, $n=500,1000,2500,5000,10000$.}
\label{fig:histteststat2}
\end{figure}

\newpage

\subsection*{Two-stage stratified cluster sampling}

\begin{figure}[htbp]
\centering
\includegraphics[width=0.99\textwidth]{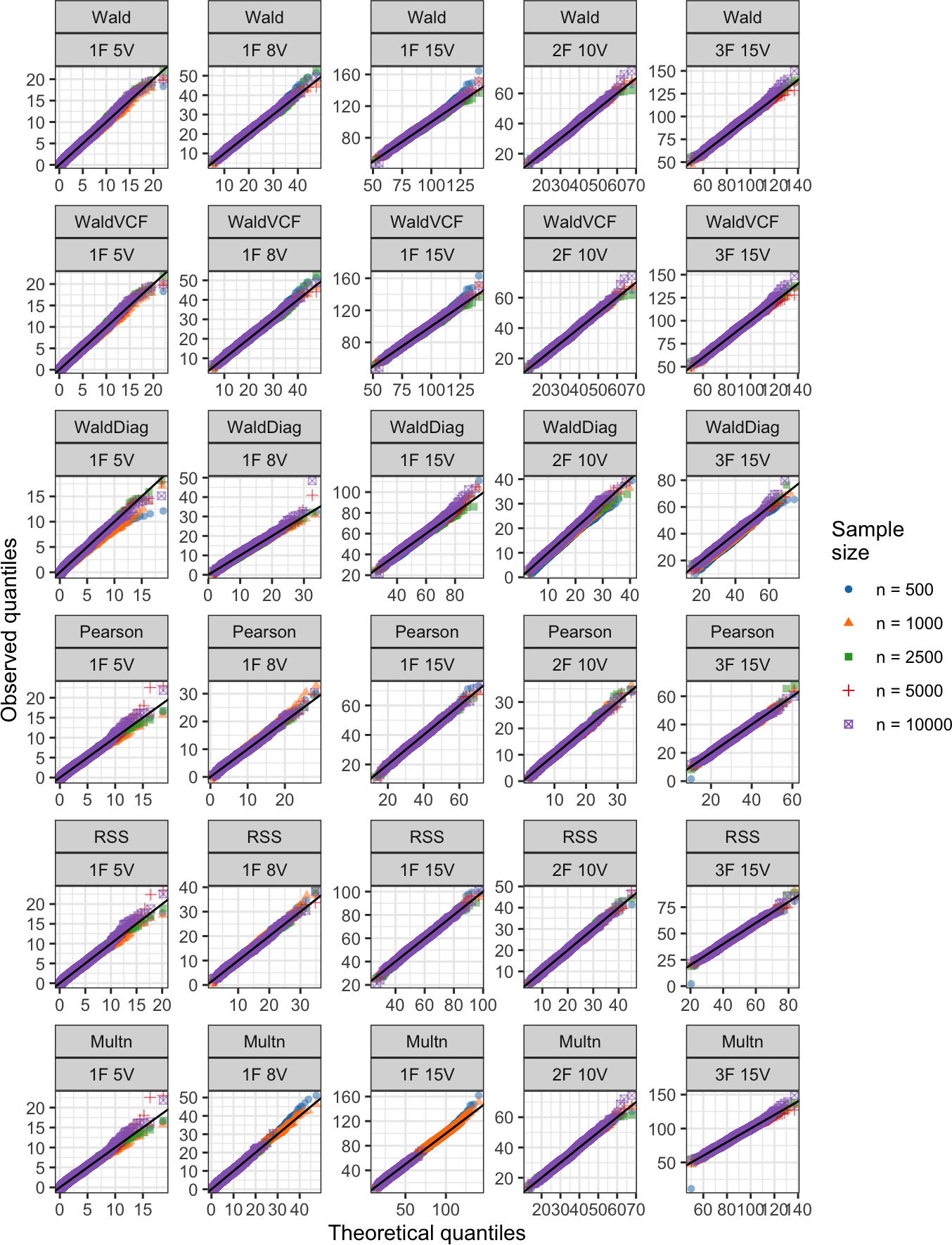}
\caption{QQ plot  for the six test statistics in Table \ref{tab:summaryteststats} across the five models in Table \ref{tab:models} from simulated data for the  two-stage stratified cluster sampling under the null hypothesis and their respective theoretical asymptotic distributions, $n=500,1000,2500,5000,10000$.}
\label{fig:histteststat3}
\end{figure}

\end{document}